\documentclass[final,onefignum,onetabnum]{siamart171218}
\usepackage[hmargin=1in,vmargin=.875in]{geometry}
\usepackage{amsmath}
\usepackage{graphicx}
\usepackage{comment}
\usepackage{wasysym}
\usepackage{placeins}
\usepackage{cases}
\usepackage{gensymb}
\usepackage[english]{babel}
\usepackage[utf8]{inputenc}
\usepackage{amsmath}
\usepackage{scalerel}
\usepackage{xcolor,cancel}
\usepackage{mathtools}
\usepackage{graphicx, caption, subcaption}
\usepackage{mdframed}
\usepackage[colorinlistoftodos]{todonotes}
\usepackage{amssymb}
\usepackage{marginnote}
\usepackage{systeme}
\usepackage{kantlipsum}
\usepackage{pdflscape}
\usepackage{soul}
\usepackage{scalerel,stackengine}
\stackMath
\usepackage{nomencl}

\makenomenclature
\usepackage{etoolbox}
\renewcommand\nomgroup[1]{%
  \item[\bfseries
  \ifstrequal{#1}{A}{Subscripts}{
  \ifstrequal{#1}{B}{Dimensional quantities}{%
  \ifstrequal{#1}{C}{Dimensionless parameters}{%
  \ifstrequal{#1}{D}{Dimensionless variables}{%
  \ifstrequal{#1}{E}{Variations on the Rayleigh Number}{%
  \ifstrequal{#1}{F}{Domains and operators}{%
  }}}}}}%
]}

\newcommand{\parD}[2]{\frac{\partial #1}{\partial #2}}
\newcommand{\DparD}[2]{\dfrac{\partial #1}{\partial #2}}

\newcommand{\vup}{\mathbf{v}}

\newcommand{\vuf}{ \mathbf{u_f} }

\newcommand{\vufp}{ \mathbf{v_f} }

\newcommand{\vum}{ \mathbf{u_m} }

\newcommand{\vump}{ \mathbf{v_m} }

\newcommand{\crn}{ R_{\textrm{E}} }

\newcommand{\gamF}{ \Gamma_{f}}
\newcommand{\gamM}{ \Gamma_{m}}

\newcommand{\alBJSJ}{\alpha}

\newcommand{\matt}[1]{{\color{black} #1}}
\setstcolor{red}

\newcommand{\Ram}{\textrm{Ra}_{m}}
\newcommand{\Raf}{\textrm{Ra}_{f}}
\newcommand{\Ramlam}{\textrm{Ra}_{m,\lambda}}
\newcommand{\Raflam}{\textrm{Ra}_{f,\lambda}}
\newcommand{\Ramax}{\textrm{Ra}_{m}}
\newcommand{\Ramc}{\textrm{Ra}_{m,c}}

\newcommand{\Ramcu}{\textrm{Ra}_{m}^*}
\newcommand{\Rafcu}{\textrm{Ra}_{f}^*}
\newcommand{\dhatc}{\hat{d}^*}

\def\reals{\hbox{\rm I\kern-.18em R}}
\def\complexes{\hbox{\rm C\kern-.43em
\vrule depth 0ex height 1.4ex width .05em\kern.41em}}
\def\field{\hbox{\rm I\kern-.18em F}} 
\usepackage{lipsum}
\usepackage{amsfonts}
\usepackage{graphicx}
\usepackage{epstopdf}
\usepackage{algorithmic}
\ifpdf
  \DeclareGraphicsExtensions{.eps,.png.pdf}
\else
  \DeclareGraphicsExtensions{.eps}
\fi

\newsiamremark{remark}{Remark}
\newsiamremark{hypothesis}{Hypothesis}
\crefname{hypothesis}{Hypothesis}{Hypotheses}
\newsiamthm{claim}{Claim}

\headers{Convection in a coupled fluid-porous media system}{M. McCurdy, M.N.J. Moore, and X. Wang}

\title{Convection in a coupled free flow-porous media system\thanks{Submitted to the editors 9 January 2019.
}}

\author{Matthew McCurdy\thanks{Department of Mathematics, Florida State University.
  (\email{mmccurdy@math.fsu.edu}, \email{mnmoore2@fsu.edu}).}
\and M.N.J.~Moore\footnotemark[2] \thanks{Geophysical Fluid Dynamics Institute, Florida State University. }
\and Xiaoming Wang\thanks{Department of Mathematics, Southern University of Science and Technology, and SUSTech International Center for Mathematics
  (\email{wxm.math@outlook.com}).}}

\usepackage{amsopn}

\begin{document}
\maketitle

\begin{abstract}\label{sect:abstr}
We perform linear and nonlinear stability analysis for thermal convection in a fluid overlying a saturated porous medium.
We use a coupled system, with the Navier-Stokes equations and Darcy's equation governing the free-flow and the porous regions respectively.
Incorporating a dynamic pressure term in the Lions interface condition (which specifies the normal force balance across the fluid-medium interface) permits an energy bound on the typically uncooperative nonlinear advection term, enabling new nonlinear stability results.
\matt{Within certain regimes, the nonlinear stability thresholds agree closely with the linear ones, and we quantify the differences that exist.}
We then compare stability thresholds produced by several common variants of the tangential interface conditions, using both numerics and asymptotics in the small Darcy number limit.
Finally, we investigate the transition between full convection and fluid-dominated convection
\matt{using both numerics and a heuristic theory. This heuristic theory is based on comparing the ratio of the Rayleigh number in each domain to its corresponding critical value, and it is shown to agree reasonably well with the numerics regarding how the transition depends on the depth ratio, the Darcy number, and the thermal-diffusivity ratio.}
\end{abstract}



\begin{DOI}

\end{DOI}

\section{Introduction}\label{sect:intro}
The phenomenon of fluid flowing over a porous medium has been observed, studied, and scrutinized for more than a century in a variety of settings.
Chief among these are geophysical applications, such as the mixing of surface water and groundwater \cite{boana2014,buss2009,cardenas2015, ewing1997,ewing1998}, contaminant transport and bioremediation efforts \cite{chen1994,suchomel1998}, and flow within oil reservoirs \cite{allen1984,allen1988}.
Given the urgent need to understand water resources more fully, investigating the interaction between surface- and groundwater is particularly timely \cite{wwap2018}.
To gain useful insight into the nature of these coupled fluid-porous systems, both linear and nonlinear stability arguments have been conducted and analyzed \cite{hill2010b, hill2010a, hill2009}.
However, the presence of  nonlinear advection $\left(\mathbf{u} \cdot \nabla\right) \mathbf{u}$ can hinder nonlinear stability analysis since, when coupled to non-trivial interface conditions, it produces a sign-indefinite term in the energy bound.
As a result, the nonlinear stability of the coupled Navier-Stokes-Darcy system---the most well-accepted model for fluid-porous systems in geophysical applications---remains unresolved.
\matt{Building upon previous work \cite{hill2010b, hill2010a,hill2009,payne1998analysis},} the primary goal of the current study is to analyze nonlinear stability of this coupled system and to examine the associated convection patterns.

To overcome the indefinite term in the energy analysis, researchers have adopted various approaches.
Several works forgo the nonlinear term altogether by exploring \textit{linear} stability of the Navier-Stokes-Darcy-Boussinesq system.
Many of these works also include additional physical effects, such as variable viscosity or permeability, quadratic equations of state for thermal expansion, and anisotropic or heterogeneous porous media \cite{carr2004,carr2003,chen1988,chen1991,chen1992, nield1977,straughan2002a}.
Other strategies to treat or avoid the nonlinear term include using Stokes in lieu of Navier-Stokes in the free-flow zone \cite{hill2010b}, or considering the Navier-Stokes-Brinkman system so that the convective term of the free-flow has a corresponding term in the porous medium \cite{hill2009}.
The Brinkman equations apply to highly porous media (e.g.~porosity greater than $.75$), which is a common and physically realistic assumption for many industrial applications such as lightweight structures, biomedical implants, heat exchangers, and chemical reactors \cite{lefebvre2008,straughan2002a}.
However, for many flows of geophysical interest (e.g.~karst aquifers, sinkholes, hyporheic flow, contaminant transport), the porosity is very small and Darcy is the most appropriate equation to model its fluid flow.

A fundamental assumption made in linear stability analyses is that perturbations to the steady-state are small and consequently, the effects of quadratic and higher order terms are lost.
As a result, there is limited information about the behavior of the nonlinear system and a possibility for subcritical instabilities---those that occur prior to the threshold predicted by the linear theory.
Nonlinear stability analyses take higher order and nonlinear terms into account, thereby providing a more holistic understanding of the mechanisms that create convection and the interplay between them.

In this work, we investigate thermal convection in a fluid overlying a saturated porous medium within the Navier-Stokes-Darcy-Boussinesq model via the energy method.
To overcome the difficulty associated with the nonlinear term, we employ the \textit{Lions interface condition}, which incorporates a dynamic pressure term into the normal-force balance\footnote{A formal asymptotic analysis justifying the smallness of this dynamic pressure at the physically important small Darcy number regime is included in Appendix II.}.
When the Lions interface condition is used in tandem with the \matt{Beavers-Joseph-Saffman-Jones (BJSJ)} condition, the Navier-Stokes system satisfies an energy law. \matt{That is, the energy associated with the nonlinear term of Navier-Stokes can be bounded.}
We outline the linear argument for the coupled system and then conduct the nonlinear stability analysis, followed by a comparison of marginal stability curves produced by each approach.
In addition, while a considerable amount of effort has been placed on determining the appropriate models for fluid flow in surface- and groundwater regions, there is less of a consensus on choosing a condition for the shear-stress balance.
Many works specify that the shear stress must balance with a jump in tangential velocity, or some variant thereof.
Popular choices for this interface condition are the Beavers-Joseph condition (BJ), the Beavers-Joseph-Jones condition (BJJ), and the BJSJ condition.
\matt{We show the relative difference between curves produced by the BJSJ condition versus those produced by either BJ or BJJ scales like the Darcy number, \matt{$\textrm{Da}$,} while the absolute differences scale like $\textrm{Da}^2$. Thus, differences between these choices are small in the physically relevant regime of small Darcy number.}

Convection in a fluid \matt{overlying} porous media is much more complex than its single layer counterparts, with more physical parameters affecting the heat transport.
One physically important phenomenon is the transition from full convection, where convection cells envelope the entire domain, to fluid-dominated convection, where the cells are confined to the free-flow region.
Parameters that influence this transition include the Darcy number, the ratio of free-flow to medium depth, and the ratio of the thermal diffusivities.
We propose a simple theory, based on comparing the critical Rayleigh numbers of the two layers, to predict this transition.
Numerical tests confirm that this theory indeed predicts the transition with \matt{reasonable} accuracy.

The rest of the article is organized as follows.
We introduce the mathematical formulation of the problem, including the governing equations, the boundary and interface conditions, and the nondimensionalization, in section 2.
We summarize the linear stability analysis in section 3 while section 4 is devoted to the nonlinear stability analysis.
Main results are outlined and discussed in section 5.
We offer our conclusions in section 6.
Nomenclature and the formal small Darcy number asymptotic expansions are included in the appendices.

\section{Formulation of the problem}\label{sect:formProb}

In this section, we describe the governing equations along with the boundary and interface conditions.
We then find steady-state solutions, which serve as reference states for the stability analyses, and we nondimensionalize the resulting system.

\subsection{Governing equations}
In the free-flow zone, we use the incompressible Navier-Stokes equations with constant viscosity and the Boussinesq approximation, coupled with the advection-diffusion equations for heat:

    \begin{align}\setlength\arraycolsep{1pt}\renewcommand\arraystretch{2}
        \left\{
        \begin{array}{rl}
        \rho_0\left( \DparD{\vuf}{t} + \left(\vuf \cdot \nabla\right) \vuf \right) &= \nabla \cdot\mathbb{T}\left(\vuf, p_f \right) - g\rho_0 \left[1-\beta\left(T_f-T_0 \right) \right]\mathbf{k}\, , \\
       \nabla \cdot \vuf &=0\, , \\
        \DparD{T_f}{t} + \vuf \cdot \nabla T_f &=\dfrac{\kappa_f}{ \left(\rho_0c_p\right)_f} \nabla^2T_f\, ,
        \end{array}\right.
    \end{align}
where $\vuf = (u_f, v_f, w_f)$, $p_f$, and $T_f$ are the free flow velocity, pressure, and temperature, respectively, with $g$, $\rho_0$, $\beta$, and $T_0$ as acceleration due to gravity, the reference density of the fluid, the coefficient of thermal expansion, and the temperature of the conductive state at the interface, respectively.
The stress tensor and rate of strain tensor are defined as $\mathbb{T}(\vuf, p_f)=2\mu_0\mathbb{D}(\vuf)-p_f\mathbb{I}$ and $\mathbb{D}(\vuf)=\frac{1}{2}\left(\nabla\vuf + \nabla \vuf^\mathsf{T}\right)$, respectively, with $\mu_0$ as dynamic viscosity and $\mathbf{k}$ as the upward pointing unit normal.
Additionally, $\kappa_f$, $c_p$, and $\lambda_f = \kappa_f/\left(\rho_0c_p\right)_f$ are the thermal conductivity of the fluid, specific heat capacity of the fluid, and thermal diffusivity of the fluid, respectively.

For fluid flow in porous media, the Darcy or Brinkman equations are the prevailing choice in the literature.
For porous media with relatively large porosity ($\chi >.75$), Brinkman is more appropriate than Darcy.
Darcy is valid under the assumption that the medium has a small porosity \cite{bear1972,nield2017}, generally applicable to geophysical systems.
We therefore employ the Darcy system with the advection-diffusion equation for heat:
    \begin{align}\setlength\arraycolsep{1pt}\renewcommand\arraystretch{2}
        \left\{
        \begin{array}{rl}
        \dfrac{\rho_0}{\chi}\DparD{\vum}{t} + \dfrac{\mu_0}{\Pi} \vum &= -\nabla p_m - g\rho_0 \left[1-\beta\left(T_m-T_L \right) \right] \mathbf{k}\, , \\
       \nabla \cdot \vum &=0\, ,\\
        \dfrac{\left(\rho_0c_p\right)_m}{ \left(\rho_0c_p\right)_f}\DparD{T_m}{t} + \vum \cdot \nabla T_m &=\dfrac{\kappa_m}{ \left(\rho_0c_p\right)_f} \nabla^2T_m\, ,
        \end{array}\label{eq:DarcySys}\right.
    \end{align}
where $\vum = (u_m, v_m, w_m),$ $p_m,$ and $T_m$ are the velocity, pressure, and temperature in the porous medium respectively, $\chi$ and $\Pi$ are the porosity and permeability, $\lambda_m = \kappa_m/\left(\rho_0c_p\right)_f$ is the thermal diffusivity of the medium, and $T_L$ is the temperature at the lower boundary of the domain.
In this work, we assume the medium to be homogeneous and isotropic so that the permeability $\Pi$ is constant and scalar-valued. For anisotropic media, \matt{the permeability $\Pi$ would be tensor-valued.}
\matt{The thermal conductivity $\kappa_m$ and specific heat capacity $(\rho_0\,c_p)_m$ of the porous medium are defined as averages of the fluid and solid components.
Many references simply use an arithmetic average \cite{hill2010b,hill2010a,hill2009}
$\phi_m = \chi \phi_f + \left(1-\chi\right)\phi_s$, where $\phi$ represents either thermal conductivity or heat capacity. However, we point out that homogenization theory gives the harmonic average, $\phi_m^{-1} = \chi \phi_f^{-1} + \left(1-\chi\right)\phi_s^{-1}$.
Though we advocate the latter approach, the analysis presented here is independent of which average is used. We also remark that, since we are studying the onset of convection, the thermal conductivity and specific heat are intrinsic values and not effective values which incorporate dispersive effects.}


The time derivative $\partial_t \vum$ in the first equation of \eqref{eq:DarcySys}  is often neglected since it is heuristically small at small Darcy number.
\matt{Inclusion of this term in Darcy's equation has been debated in the literature \cite{vafai2005handbook}. In this paper, we include the time derivative primarily for the benefit of the energy analysis, although a welcome side effect is that this term would allow more accurate description of temporal transitions.}
Several works concerning linear stability, \cite{ vertlayer2012,chen1991,chen1992,nield1996}, exclude time derivatives of the Navier-Stokes and/or Darcy equations by invoking the principle of exchange of stabilities. \matt{We note that this principle has not been rigorously established for the coupled system, and therefore we do not assume it. Our numerics, however, suggest that the principle seems to hold in practice.}

\begin{figure}[!ht]
  \centering
{\includegraphics[width=0.45\textwidth]{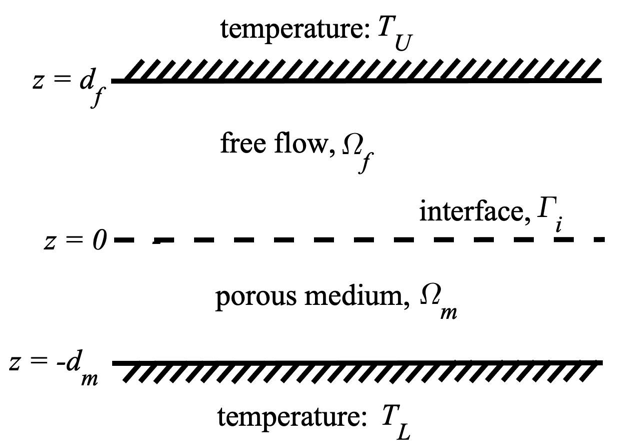}}
  \caption{Schematic of the domain $\Omega = \{(x,y)\in\mathbb{R}^2 \times z\in(-d_m,d_f)\}$, comprised of a free-flow region $\Omega_f$ and a porous medium $\Omega_m$. The two subdomains meet at an interface $\Gamma_i$. The upper and lower boundaries are impermeable and held at constant temperatures $T_U$ and $T_L$, respectively, with $T_L>T_U.$\label{fig:dom}}
\end{figure}

\subsection{Boundary and interface conditions}
For the domain, shown in Figure \ref{fig:dom}, we assume flat, horizontal, non-penetrable plates at the top and bottom with a non-deforming interface between the two regions, $\Omega_f=\{(x,y)\in \mathbb{R}^2 \times z\in (0,d_f)\}$ for the free flow and $\Omega_m=\{(x,y)\in \mathbb{R}^2 \times z\in (-d_m,0)\}$ for the porous medium.
The temperature is held constant at the top and bottom plates.
For the flow, we use a free-slip condition at the top and an impermeable condition at the bottom,
    {
    \begin{align}\setlength\arraycolsep{1pt}\renewcommand\arraystretch{1.25}
        \left\{
        \begin{array}{cccccccl}
        T_f &= T_U\, , &\qquad  &\vuf \cdot\textbf{n}= \parD{\vuf_\tau}{\textbf{n}} &= 0\, ,& \qquad  &\textrm{at }z&=d_f\, ,\\
        T_m &= T_L\, , &  &\vum \cdot \textbf{n} &= 0\, ,& \qquad &\textrm{at }z&=-d_m\, ,\\
        \end{array}\right.
    \end{align}
    }
where ${\vuf}_\tau=(v_f, w_f)$ denotes the tangential (horizontal) components of the velocity at the top of the domain with $\textbf{n}$ as the unit normal vector.

At the interface $\Gamma_i$ $(z=0)$, we require continuity of temperature, heat flux, and the normal component of velocity:
    \begin{align}
        T_f &= T_m\, ,\\
        k_f\,\nabla T_f \cdot \mathbf{n} &=  k_m\,\nabla T_m \cdot \mathbf{n}\, ,\\
         \vuf \cdot \mathbf{n} &=  \vum \cdot \mathbf{n}\, .
    \end{align}

The next condition must involve the tangential stress at the interface. The primary condition considered here is the Beavers-Joseph-Saffman-Jones (BJSJ) condition \cite{saffman1971}, also known as the Navier-slip condition, which relates the shear stress to the tangential velocity:
    \begin{align}\label{eq:BJSJcond}
         -\boldsymbol{\tau} \cdot \mathbb{T} \left(\vuf,p_f \right)\mathbf{n} 
         &= \frac{\mu_0\,\alBJSJ}{\sqrt{\Pi}}\boldsymbol{\tau} \cdot \vuf \, ,
    \end{align}
where $\alBJSJ$ is an empirically determined coefficient and $\boldsymbol{\tau}$ denotes the unit tangent vectors. Our nonlinear stability analysis will rely on the BJSJ condition.
This condition is debated in the literature, and we will therefore consider a few alternatives in the linear analysis, namely the Beavers-Joseph (BJ) condition \cite{beavers1967} and the Beavers-Joseph-Jones (BJJ) condition \cite{jones1973}.
All three conditions can be represented concisely as
    \begin{align*}
    \parD{u_{f,\gamma}}{z} + \Psi_{J}\,\parD{w_{f}}{x_\gamma} = \frac{\alBJSJ}{\sqrt{\Pi}}\left(u_{f,\gamma}- \Psi_{S}\,u_{m,\gamma} \right) \quad \textrm{for }\gamma=1,2 \, ,
    \end{align*}
where $u_{i,\gamma}$ is the $\gamma$ component of the velocity in $\Omega_i$, and $\Psi_{J},\Psi_{S}\in \{0,1\}$ are switches associated with the Jones correction and the Saffman approximation terms, respectively.
The BJ condition corresponds to $\Psi_{J}=0$ and $\Psi_{S}=1$, the BJJ condition to $\Psi_{J}=1$ and $\Psi_{S}=1$, and the BJSJ condition to $\Psi_{J}=1$ and $\Psi_{S}=0$.
\matt{The BJJ is considered to be the most physically accurate, as it relates the shear stress to the jump in the tangential velocity across the interface. The BJ condition omits the term ${\partial w_{f}}/{\partial x_\gamma }$ in the shear stress  \cite{carr2003,chen1988,chen1991,chen1992}. Meanwhile, the Saffman approximation drops the Darcy velocity in the right-hand-side, which is relatively small in magnitude as long as the Darcy number is small. Thus, since our nonlinear analysis relies on the BJSJ condition, it will be limited to the physically relevant regime of small Darcy number.}

In \cite{straughan2002a}, Straughan compares the BJ and BJJ conditions, showing that the linear marginal stability curves produced by each are almost the same.
In section \ref{sect:results}, we expand upon Straughan's findings by showing that the three interface conditions each produce similar marginal stability curves.
Specifically, we show the relative difference between curves produced by the BJSJ condition versus those produced by either BJ or BJJ scales like $\textrm{Da}$ in the small Darcy number regime.

The last interface condition concerns the balance of force in the normal direction, and there are two options:
    \begin{align}
    \label{normal_cond}
    -\mathbf{n} \cdot \mathbb{T}\left(\vuf,p_f \right)\mathbf{n} +\Psi_{L}\,\frac{\rho_0}{2}\left|\vuf \right|^2 = p_m\, ,
    \end{align}
where $\Psi_{L}\in \{0,1\}$ is a switch for the dynamic pressure term, $\frac{\rho_0}{2}|\vuf|^2$.
The most common choice in the literature is $\Psi_{L}=0$ which renders Eq.~\eqref{normal_cond} linear.
\matt{ For nonlinear analysis, we will choose $\Psi_{L}=1$, which is known as the {\em Lions interface condition} \cite{ccecsmeliouglu2008analysis, ccecsmeliouglu2009primal, chidyagwai2007weak, discacciati2009navier,girault2009dg}.
This choice gives rise to an energy law which facilitates the analysis significantly. In the appendix, we show that the dynamic pressure term is order $\textrm{Da}$ for small Darcy number.}

\subsection{Steady-state and perturbed system}
First, we introduce the following steady-state solution, known as the conductive state (denoted with an overhead bar):
    \begin{align*}
    \bar{\mathbf{u}}_{\mathbf{f}} &= \bar{\mathbf{u}}_{\mathbf{m}} = 0\, , \\
     \bar{T}_f&= T_0 + z\frac{T_U - T_0 }{d_f}\, ,\\
     \bar{T}_m &= T_0 + z\frac{T_0 - T_L }{d_m} \, .
    \end{align*}
Here, $T_0$ represents the interface temperature of the conductive solution
    \begin{align*}
    T_0 = \frac{\kappa_m\, d_f\, T_L + \kappa_f \,d_m\, T_U}{\kappa_m \,d_f + \kappa_f \,d_m}\, .
    \end{align*}
If $T_U>T_L$, the conductive state is stable, but if $T_L>T_U$, buoyancy can destabilize the system. In this paper, we consider the latter case.
Additionally, we choose $\bar{p}_f$ and $\bar{p}_m$ to satisfy
    \begin{align*}
    \nabla \bar{p}_f &= -g\rho_0\left(1-\beta\left(\bar{T}_f-T_0\right)\right)\mathbf{k}\, , \\
    \nabla \bar{p}_m &= -g\rho_0\left(1-\beta\left(\bar{T}_m-T_L\right)\right)\mathbf{k}\, .
    \end{align*}
We perturb the steady-state as follows:
    \begin{align} \label{eq:perturbedSolns}
    \vuf &= \bar{\mathbf{u}}_{\mathbf{f}} + \vufp\, , \quad \vum =  \bar{\mathbf{u}}_{\mathbf{m}} +\vump\, \nonumber,\\
    T_f &= \bar{T}_f +\theta_f\, , \quad T_m = \bar{T}_m+\theta_m\, ,  \\
    p_f &= \bar{p}_f +\pi_f\, , \quad p_m= \bar{p}_m + \pi_m \, ,  \nonumber
    \end{align}
where $\textbf{v}_j,$ $\theta_j,$ and $\pi_j$ are the perturbation variables.
In the linear stability analysis, the perturbations are assumed to be small compared to the background state.
However, with the nonlinear analysis, there is no assumption concerning the magnitude of the perturbations.
Substituting \eqref{eq:perturbedSolns} into the original system produces:
     \begin{equation*}
     \textrm{In }\Omega_f: \quad \begin{dcases}
    &\rho_0 \left(\parD{\vufp}{t} +\left(\vufp\cdot \nabla\right) \vufp \right)= \nabla \cdot\mathbb{T}\left(\vufp, \pi_f \right) + \rho_0 g \beta \theta_f \mathbf{k}\, ,\\
    &\nabla \cdot \vufp = 0\, ,\\
    &\parD{\theta_f}{t}+ \vufp \cdot \nabla \theta_f = \lambda_f \nabla^2 \theta_f - w_f \left(\frac{T_U-T_0}{d_f}\right)\, ,
    \end{dcases}
    \end{equation*}
for $(x,y,z,t)\in \{\mathbb{R}^2\times (0,d_f)\times (0,\infty)\}$,
    \begin{equation*}
     \textrm{In }\Omega_m: \quad\begin{dcases}
    &\frac{\rho_0}{\chi} \parD{\vump}{t} + \frac{\mu_0}{\Pi}\vump= -\nabla \pi_m  + \rho_0 g \beta \theta_m \mathbf{k}\, ,\\
    &\nabla \cdot \vump = 0\, ,\\
    &\varrho\, \parD{\theta_m}{t}+\vump\cdot \nabla \theta_m = \lambda_m \nabla^2 \theta_m -w_m \left(\frac{T_0-T_L}{d_m}\right)\, , \end{dcases}
    \end{equation*}
for $(x,y,z,t)\in \{\mathbb{R}^2\times (-d_m,0)\times (0,\infty)\}$, and
    \begin{equation*}
     \textrm{On }\Gamma_i: \quad\begin{dcases}
    &\theta_f = \theta_m\, ,\\
    &\kappa_f \nabla \theta_f\cdot \mathbf{n}=\kappa_m \nabla \theta_m\cdot \mathbf{n}\, ,\\
    &\vufp \cdot \mathbf{n} = \vump \cdot \mathbf{n}\, , \\
    &-\boldsymbol{\tau} \cdot \mathbb{T}\left(\vufp, \pi_f \right)\mathbf{n} = \frac{\mu_0\,\alBJSJ}{\sqrt{\Pi}}\left(\boldsymbol{\tau} \cdot \vufp \right)\, ,\\
    &-\mathbf{n}\cdot\mathbb{T}\left(\vufp, \pi_f \right)\mathbf{n}+\Psi_{L}\,\frac{\rho_0}{2}|\vufp|^2 = \pi_m\, ,
    \end{dcases}
    \end{equation*}
for $(x,y,0,t)\in \{\mathbb{R}^2\times (z=0)\times (0,\infty)\}$ with
$$\varrho = \frac{\left(\rho_0c_p\right)_m}{ \left(\rho_0c_p\right)_f}\, .$$

\subsection{Nondimensionalization}

We introduce the same scalings as \cite{chen1988,straughan2002a} with nondimensional variables denoted by tildes:
    \begin{alignat*}{5}
    & \vufp = \tilde{\textbf{v}}_f \frac{\nu}{d_f}\, ,  \qquad && \textbf{x}_f = \tilde{\textbf{x}}_f \,d_f\, , \qquad        && t_f = \tilde{t}_f\, \frac{d_f^2}{\lambda_f}\, , \qquad &&\theta_f = \tilde{\theta}_f \,\frac{\left(T_0-T_U\right)\nu}{\lambda_f}\, ,\qquad         && \pi_f = \tilde{\pi}_f \,\frac{\rho_0\, \nu^2}{d_f^2}\, ,\\
    & \vump = \tilde{\textbf{v}}_m \frac{\nu}{d_m}\, , \qquad && \textbf{x}_m = \tilde{\textbf{x}}_m \,d_m\, , \qquad    && t_m = \tilde{t}_m\, \frac{d_m^2}{\lambda_m}\, , \qquad && \theta_m = \tilde{\theta}_m \,\frac{\left(T_L-T_0\right)\nu}{\lambda_m}\, ,    \qquad    && \pi_m = \tilde{\pi}_m \,\frac{\rho_0\, \nu^2}{d_m^2}\, ,
    \end{alignat*}
(where $\nu = \mu_0/\rho_0$ is the kinematic viscosity) which yields the systems (sans tildes):
    \begin{equation} \label{nondimcoupFree}
    \textrm{In }\Omega_f: \quad\begin{dcases}
    &\frac{1}{\textrm{Pr}_f}\parD{\vufp}{t_f} +\left(\vufp\cdot \nabla\right) \vufp = 2\nabla \cdot\mathbb{D}\left(\vufp\right)-\nabla \pi_f - \Raf\, \theta_f \mathbf{k}\, ,\\
    &\nabla \cdot \vufp = 0\, ,\\
    &\parD{\theta_f}{t}+ \textrm{Pr}_f\,\vufp \cdot \nabla\theta_f = \nabla^2 \theta_f - w_f\, ,
    \end{dcases}
    \end{equation}
for $(x,y,z,t)\in \{\mathbb{R}^2\times (0,1)\times (0,\infty)\}$,
    \begin{equation}\label{nondimcoupPor}
    \textrm{In }\Omega_m: \quad
    \begin{dcases}
    &\frac{1}{\chi}\frac{\textrm{Da}}{\textrm{Pr}_m}\, \parD{\vump}{t_m} + \vump = -\textrm{Da}\,\nabla{\pi_m} - \Ram\, \theta_m \mathbf{k}\, ,\\
    &\nabla \cdot \vump = 0\, ,\\
    &\varrho\, \parD{\theta_m}{t}+\textrm{Pr}_m\,\vump\cdot \nabla \theta_m =\nabla^2 \theta_m -w_m\, ,
    \end{dcases}
    \end{equation}
for $(x,y,z,t)\in \{\mathbb{R}^2\times (-1,0)\times (0,\infty)\}$, and
    \begin{equation}\label{nondimcoupInter}
    \textrm{On }\Gamma_i: \quad
    \begin{dcases}
    &\hat{d}\theta_f = \epsilon_T^2\,\theta_m\, ,\\
    & \nabla_f \theta_f\cdot \mathbf{n}= \epsilon_T \nabla_m \theta_m\cdot \mathbf{n}\, ,\\
    &\vufp \cdot \mathbf{n} = \hat{d}\, \vump \cdot \mathbf{n}\, , \\
    &-\,\boldsymbol{\tau} \cdot \mathbb{T}\left(\vufp, \pi_f \right)\mathbf{n} = \frac{\hat{d}\,\alBJSJ}{\sqrt{\textrm{Da}}}\,\left(\boldsymbol{\tau} \cdot \vufp \right)\, ,\\
    &-\mathbf{n}\cdot\mathbb{T}\left(\vufp, \pi_f\right)\mathbf{n}+\Psi_{L}\,\frac{1}{2}|\vufp|^2 = \hat{d}^2\,\pi_m\, ,
    \end{dcases}
    \end{equation}
for $(x,y,0,t)\in \{\mathbb{R}^2\times (z=0)\times (0,\infty)\}$.
Here, the notation $\nabla_j$ indicates the gradient with respect to $\textbf{x}_j$ where $j\in\{f,m\}$. 

We have introduced a total of seven dimensionless parameters.
The first five are given by
    \begin{align*}
    \hat{d}=\frac{d_f}{d_m}\, , \quad \epsilon_T = \frac{\lambda_f}{\lambda_m}\, , \quad \textrm{Da} = \frac{\Pi}{d_m^2}\, , \quad
    \textrm{Pr}_f = \frac{\nu}{\lambda_f}\, , \quad \textrm{Pr}_m = \frac{\nu}{\lambda_m}\, .
    \end{align*}
These parameters are, respectively, the depth ratio, the ratio of thermal diffusivities, the Darcy number, and the Prandtl numbers of the free-flow and porous regions.
The last two are the Rayleigh numbers of the two regions
    \begin{align}\label{eq:RafRamrelation}
    \Raf = \frac{g\beta \left(T_0-T_U \right) d_f^3}{\nu \,\lambda_f}\, , \quad  \Ram = \frac{g\beta \left(T_L-T_0 \right) \textrm{Da}\, d_m^3}{\nu \,\lambda_m} = \Raf \frac{\textrm{Da}\,\epsilon_T^2}{\hat{d}^4}\, .
    \end{align}

\section{Linear Stability}\label{sect:LSA}
In this section, we briefly overview the linear stability analysis of system \eqref{nondimcoupFree}--\eqref{nondimcoupInter}.
For additional details, the reader is  referred to \cite{straughan2002a}, which differs only in the interface condition chosen in Eq.~\eqref{eq:BJSJcond}.
Here, we set $\Psi_{J} = 1,$ $\Psi_{S} = 0$, corresponding to the BJSJ condition.
The value of $\Psi_{L}$ is irrelevant since the dynamic pressure term is nonlinear and hence omitted in linear analysis.

Assuming perturbations to be small eliminates quadratic and higher-order terms from \eqref{nondimcoupFree}--\eqref{nondimcoupInter}.
With the resulting linear system, we take the double curl to remove the pressure terms and then, considering the third component, we substitute normal mode solutions
    \begin{align} \label{eq:NormModeSol}
    w_j(\mathbf{x},t) = F_j(x,y)\, \tilde{w}_j(z)e^{\sigma_j t}\,  \quad\textrm{and} \quad \theta_j (\mathbf{x},t) = F_j(x,y)\, \tilde{\theta}_j(z) e^{\sigma_j t}\, ,
    \end{align}
    for $j \in\{f,m\}$. 
    Here, $F_j(x,y)$ corresponds to a unimodal component the horizontal planform in each region with corresponding horizontal wavenumber $a_j$. That is,
    \begin{align*}
   a_j^2\,F_j(x,y) + \nabla_H^2 \,F_j(x,y)=0\,
    \end{align*}
where $\nabla_H^2 = \parD{^2}{x^2}+\parD{^2}{y^2}$ is the horizontal Laplacian operator. This modal decomposition defines the structure of the convection cells \cite{carr2003,drazin2004hydrodynamic, straughan2004resonant,straughan2013energy}.
 With \eqref{eq:NormModeSol}, the real part of $\sigma_j$ determines the stability of the flow; if $\text{Re}(\sigma_j) <0$ the corresponding normal mode decays in time and if $\text{Re}(\sigma_j) >0$ it grows.
From our nondimensional scalings, we note the following relationships:
    \begin{align} \label{eq:relation}
    a_f = \hat{d}\,a_m\, ,\quad \sigma_f = \frac{\epsilon_T}{\hat{d}^2}\sigma_m\, , \quad \Raf = \Ram \frac{\hat{d}^4}{\textrm{Da}\,\epsilon_T^2}\, .
    \end{align}

Using the notation $D_f = \frac{d}{dz_f}$ and $D_m = \frac{d}{dz_m}$ for spatial derivatives in $\Omega_f$ and $\Omega_m$, respectively, we acquire the system:
    \begin{equation} \label{eq:nonDlinTop}
     \textrm{In }\Omega_f,\,\, z\in (0,1): \quad\begin{dcases}
    &\frac{\sigma_f}{\textrm{Pr}_f} {\matt{\left( D_f^2-a_f^2\right)w_f}} ={\matt{\left(D_f^2 - a_f^2 \right)^2w_f}}  + a_f^2 \,\Raf\,\theta_f\, ,\\
    &\sigma_f \theta_f =\left(D_f^2 - a_f^2 \right)\theta_f - w_f\, ,
    \end{dcases}
    \end{equation}

    \begin{equation}\label{eq:nonDlinBot}
    \textrm{In }\Omega_m,\,\, z\in (-1,0): \quad
    \begin{dcases}
    &\frac{\sigma_m}{\chi}\frac{\textrm{Da}}{\textrm{Pr}_m}\left(D_m^2 - a_m^2 \right) w_m = -\left(D_m^2 - a_m^2 \right)w_m  +a_m^2\, \Ram\, \theta_m \, ,\\
    &\sigma_m\,\varrho\, \theta_m = \left(D_m^2 - a_m^2 \right)\theta_m - w_m \, ,
    \end{dcases}
    \end{equation}

    \begin{equation}\label{eq:nonDlinIC}
    \textrm{On }\Gamma_i, \,\, z = 0: \quad
    \begin{dcases}
    &\hat{d}\theta_f = \epsilon_T^2\,\theta_m\, ,\\
    & D_f\theta_f= \epsilon_T D_m\theta_m\, ,\\
    &w_f = \hat{d}\, w_m\, , \\
    &D_f^2w_f  = \frac{\hat{d}\,\alBJSJ}{\sqrt{\textrm{Da}}}D_fw_f\, ,\\
    &\frac{\hat{d}^4}{\chi \textrm{Pr}_m}\sigma_m\,D_mw_m + \frac{\hat{d}^4}{\textrm{Da}}D_m w_m = \frac{\sigma_f}{\textrm{Pr}_f}D_fw_f - D_f^3w_f +3a_f^2D_fw_f\, ,
    \end{dcases}
    \end{equation}
with the boundary conditions at the top and bottom of the domain:
    \begin{align}
    &\textrm{At  }z = 1: \quad w_f = D_fw_f = \theta_f = 0\, ,\\
    &\textrm{At  }z = -1: \quad w_m = \theta_m = 0\, .\label{eq:nonDlinBC}
     \end{align}

System \eqref{eq:nonDlinTop}--\eqref{eq:nonDlinBC} constitutes a generalized eigenvalue problem for either $\sigma_f$ or $\sigma_m$, which we solve with the Chebyshev tau-QZ algorithm \cite{dongarra1996} implemented with the Chebfun package \cite{driscoll2014}.
This algorithm first performs Chebyshev collocation \cite{kopriva1998staggered, fastcheby2017, Trefethen2013} and then solves the resulting linear system with the QZ method \cite{golub2012matrix,moler1973algorithm}.
Lastly, we make substitution \eqref{eq:relation} to find the marginal stability curves in the $(a_m, \Ram)$ plane.
For each wavenumber $a_m$, there is a Rayleigh number $\Ram$ where the flow transitions from stable to unstable (i.e. {$\textrm{Re}(\sigma_j)$ changes from negative to positive).
The marginal stability curves, $\textrm{Re}(\sigma_j) = 0$, shown in section \ref{sect:results} delineate the boundary between stable and unstable regimes.

\section{Nonlinear Stability}\label{sect:NLSA}

\matt{In this section, we address nonlinear stability using the energy method.
Our analysis builds off of important previous works \cite{hill2010b, hill2010a,hill2009,payne1998analysis} that examined nonlinear stability of related fluid-porous systems. Here, we adopt similar techniques, in conjunction with the Lions interface condition, to obtain an energy law and ultimately resolve nonlinear stability of the Navier-Stokes-Darcy system.}

Throughout this section we employ the BJSJ condition ($\Psi_{J}=1, \Psi_{S}=0$) and the Lions condition ($\Psi_{L}=1$).
We use the following notation for vector-valued functions $\textbf{f}$ and $\textbf{g}$ and matrix-valued functions $\textbf{A}$ and $\textbf{B}$:
    \begin{align*}
    \left(\textbf{f}, \textbf{g} \right)_j = \int_{\Omega_j}\textbf{f}\cdot \textbf{g}\,d\Omega_j\, , \hspace{.15in}\langle\textbf{A}, \textbf{B} \rangle_j = \int_{\Omega_j}\textbf{A}\colon \textbf{B}\,d\Omega_j\, , \hspace{.15in} \|\textbf{f}\,\|_j^2=(\textbf{f},\textbf{f}\,)_j\, , \hspace{.15in} |\textbf{f}\,|^2 = \textbf{f}\cdot \textbf{f}\, ,
    \end{align*}
for domains $j \in \{f,m\}$. In this section, $\Omega_f$ and $\Omega_m$ represent a single period cell in the respective domains.
We dot the first equation of \eqref{nondimcoupFree} with $\vufp$ and integrate over $\Omega_f$:
    \begin{align*}
    \frac{1}{\textrm{Pr}_f}\left(\parD{\vufp}{t}, \vufp\right)_f +\left( \left(\vufp\cdot \nabla\right) \vufp, \vufp\right)_f = 2\left(\nabla \cdot\mathbb{D}\left(\vufp\right), \vufp\right)_f-\left(\nabla \pi_f, \vufp\right)_f -\Raf\, \left(\theta_f \mathbf{k}, \vufp \right)_f\, .
    \end{align*}
After integrating by parts, the boundary integrals reduce to integrals along the interface of the fluid region, $\gamF$, leaving
    \begin{align*}
    \frac{1}{2\,\textrm{Pr}_f}\frac{d}{dt}\|\vufp\|^2_f
    & = \frac{1}{2}\int_{\gamF}|\vufp|^2  \left(\vufp\cdot \mathbf{n} \right)\,d\gamF
     -2\langle \mathbb{D}(\vufp), \mathbb{D}(\vufp)\rangle_f- 2\int_{\gamF}\mathbf{n}\cdot \mathbb{D}(\vufp)\mathbf{n}\left(\vufp \cdot \mathbf{n} \right)\,d\gamF \\
    & - 2\int_{\gamF}\sum_{i=1}^{2}\boldsymbol{\tau}_i\cdot \mathbb{D}(\vufp)\mathbf{n}\left(\vufp \cdot \boldsymbol{\tau}_i \right)\,d\gamF +\int_{\gamF}\pi_f\left(\vufp\cdot \mathbf{n} \right)\,d\gamF-\Raf\left(\theta_f , w_f\right)_f\, ,
    \end{align*}
where $\boldsymbol{\tau}_i$ are the unit tangents in $x$ and $y$ at the interface.
Applying the BJSJ and Lions interface conditions from \eqref{nondimcoupInter} gives
    \begin{align*}
    \frac{1}{2\,\textrm{Pr}_f}\frac{d}{dt}\|\vufp\|^2_f
    & = \frac{1}{2}\int_{\gamF}|\vufp|^2 \left(\vufp\cdot \mathbf{n} \right)\,d\gamF
    -2\langle \mathbb{D}(\vufp), \mathbb{D}(\vufp)\rangle_f+\int_{\gamF}\left[\hat{d}^2\,\pi_m - \pi_f  - \frac{1}{2}|\vufp|^2\right]\left(\vufp \cdot \mathbf{n} \right)\,d\gamF \\
    &+ \int_{\gamF}\sum_{i=1}^{2}\left[\frac{\hat{d}\,\alBJSJ}{\sqrt{\textrm{Da}}}\left(\vufp \cdot \boldsymbol{\tau}_i \right) \right]\left(\vufp \cdot \boldsymbol{\tau}_i \right)\,d\gamF
    + \int_{\gamF}\pi_f\left(\vufp\cdot \mathbf{n} \right)\,d\gamF-\Raf\left(\theta_f , w_f\right)_f \, .
    \end{align*}

We note that the first term on the RHS involving $\frac{1}{2}|\vufp|^2\left(\vufp \cdot \mathbf{n} \right)$ arises from the nonlinear advection.
Importantly, the application of the {\em Lions} interface condition to the expression $\mathbf{n}\cdot \mathbb{D}(\vufp)\mathbf{n}\left(\vufp \cdot \mathbf{n} \right)$ produces a similar term with opposite sign that cancels this first term.
Without this cancellation, the presence of the sign-indefinite term $\frac{1}{2}|\vufp|^2\left(\vufp \cdot \mathbf{n} \right)$ would hamper energy analysis. However, with the cancellation, we obtain the following energy law
    \begin{align} \label{eq:en1}
    \frac{1}{2\,\textrm{Pr}_f}\frac{d}{dt}\|\vufp\|^2_f =&-2\langle \mathbb{D}(\vufp), \mathbb{D}(\vufp)\rangle_f-\Raf\left(\theta_f , w_f\right)_f  \\
    &+\int_{\gamF}\hat{d}^2\,\pi_m \left(\vufp \cdot \mathbf{n} \right)\,d\gamF +\int_{\gamF}\sum_{i=1}^{2}\frac{\hat{d}\,\alBJSJ}{\sqrt{\textrm{Da}}}\left(\vufp \cdot \boldsymbol{\tau}_i \right)^2\,d\gamF\, .  \notag
    \end{align}

Now, we dot the third equation of \eqref{nondimcoupFree}, the first equation of \eqref{nondimcoupPor}, and the third equation of \eqref{nondimcoupPor} with $\theta_f$, $\vump$, and $\theta_m$, respectively, and then integrate over the appropriate domains, producing
    \begin{align}
    \label{eq:en2}
    \frac{1}{2}\frac{d}{dt}\|\theta_f\|_f^2 &=-\frac{\textrm{Pr}_f}{2}\int_{\gamF} \left( \vufp \cdot \textbf{n} \right)\theta_f^2\,d\gamF +\int_{\gamF} \theta_f\left(\nabla \theta_f \cdot \textbf{n}\right)\,d\gamF -\|\nabla \theta_f\|_f^2 -\left(w_f, \theta_f\right)_f\, ,\\[10pt]
    \frac{1}{2}\frac{\textrm{Da}}{\chi\,\textrm{Pr}_m}\frac{d}{dt}\|\vump\|^2_m &= -\|\vump\|_m^2 + \int_{\gamM}\textrm{Da}\,\pi_m\left( \vump\cdot\mathbf{n} \right)\,d\gamM - \Ram\left(\theta_m, w_m\right)_m\, ,\label{eq:en3}\\[10pt]
    \label{eq:en4}
    \frac{\varrho}{2}\frac{d}{dt}\|\theta_m\|_m^2 &=\frac{\textrm{Pr}_m}{2}\int_{\gamM}\left(\vump\cdot \textbf{n} \right)\theta_m^2\,d\gamM -\int_{\gamM}\theta_m\left(\nabla \theta_m \cdot \textbf{n}\right)\,d\gamM - \|\nabla \theta_m\|_m^2 -\left(w_m, \theta_m\right)_m\, ,
    \end{align}
where $\gamM$ denotes the interface of the porous medium.
From here, we follow an argument similar to that of Straughan, Carr, and Hill in \cite{hill2010a,hill2009}.
We add equations (\ref{eq:en1})-(\ref{eq:en4}) together and multiply (\ref{eq:en2}), (\ref{eq:en3}), (\ref{eq:en4}) by coupling parameters $\lambda_1, \lambda_2, \lambda_3>0,$ respectively.
The introduction of these parameters permits sharper bounds on the critical Rayleigh numbers than could be obtained otherwise.
In addition, we rescale time derivatives in the porous medium by the factor $\epsilon_T/\hat{d}^2$, so that we are using the same scale as in the free-flow zone.
These manipulations yield the system
    \begin{align*}
    &\frac{d}{dt}\left[\frac{1}{2\,\textrm{Pr}_f}\|\vufp\|^2_f+\frac{\lambda_2}{2}\frac{\epsilon_T}{\hat{d}^2}\frac{\textrm{Da}}{\chi\,\textrm{Pr}_m}\|\vump\|^2_m + \frac{\lambda_1}{2}\|\theta_f\|_f^2+\frac{\lambda_3}{2}\frac{\varrho\,\epsilon_T}{\hat{d}^2}\|\theta_m\|_m^2 \right]=\\
    &\qquad\qquad -2\langle \mathbb{D}(\vufp), \mathbb{D}(\vufp)\rangle_f- \int_{\gamF}\hat{d}^2\,\pi_m \left(\vufp \cdot \mathbf{n} \right)\,d\gamF -\int_{\Gamma_i}\sum_{i=1}^{2}\frac{\hat{d}\,\alBJSJ}{\sqrt{\textrm{Da}}}\left(\vufp \cdot \boldsymbol{\tau}_i \right)^2\,d\Gamma_i-\Raf\,\left(\theta_f, w_f \right)_f \\
    &\qquad\qquad+\lambda_1\left(-\frac{\textrm{Pr}_f}{2}\int_{\gamF} \left( \vufp \cdot \textbf{n} \right)\theta_f^2\,d\gamF +\int_{\gamF} \theta_f\left(\nabla \theta_f \cdot \textbf{n}\right)\,d\gamF -\|\nabla \theta_f\|_f^2 -\left(w_f, \theta_f\right)_f\right)\\
    &\qquad\qquad+\lambda_2\left(-\|\vump\|_m^2 + \int_{\gamM}\textrm{Da}\,\pi_m\left( \vump\cdot\mathbf{n} \right)\,d\gamM - \Ram\,\left( \theta_m,w_m\right)_m \right)\\
    &\qquad\qquad+\lambda_3 \left(\frac{\textrm{Pr}_m}{2}\int_{\gamM}\left(\vump\cdot \textbf{n} \right)\theta_m^2\,d\gamM -\int_{\gamM}\theta_m\left(\nabla \theta_m \cdot \textbf{n}\right)\,d\gamM - \|\nabla \theta_m\|_m^2 -\left(w_m, \theta_m\right)_m\right)\, .
    \end{align*}

We will next choose the coupling parameters, $\lambda_1, \lambda_2, \lambda_3$, to make convenient cancellations with integrals along the interface.
First, focusing on $\lambda_2$, a change of variables allows us to write
    \begin{align*}
    -\int_{\gamF}\hat{d}^2\,\pi_m \left( \vufp \cdot \textbf{n}\right)\,d\gamF +\lambda_2 \int_{\gamM}\textrm{Da}\,\pi_m \left( \vump \cdot \textbf{n}\right)\,d\gamM
    =\left(-\hat{d}^2\,+ \lambda_2 \frac{\textrm{Da}}{\hat{d}} \right)\int_{\gamF} \pi_m \left( \vufp \cdot \textbf{n}\right)\,d\gamF \, .
    \end{align*}
We therefore choose $\lambda_2 = \hat{d}^3/\textrm{Da}$ so that the expression on the right-hand-side vanishes.
Next, consider the terms associated with $\lambda_1$ and $\lambda_3$:
    \begin{align*}
    -\lambda_1 \,\frac{\textrm{Pr}_f}{2}\int_{\gamF} \left( \vufp \cdot \textbf{n} \right)\theta_f^2\,d\gamF + \lambda_3\,\frac{\textrm{Pr}_m}{2} \int_{\gamM}\left(\vump\cdot \textbf{n} \right)\theta_m^2\,d\gamM
    = \left(-\lambda_1 + \lambda_3\,\frac{\hat{d}}{\epsilon_T^3}\right)\int_{\gamF} \left( \vufp \cdot \textbf{n} \right)\theta_f^2\,d\gamF \, , \\
    \lambda_1 \int_{\gamF} \theta_f\left(\nabla \theta_f \cdot \textbf{n}\right)\,d\gamF - \lambda_3 \int_{\gamM}\theta_m\left(\nabla \theta_m \cdot \textbf{n}\right)\,d\gamM
    = \left( \lambda_1- \lambda_3\frac{\hat{d}}{\epsilon_T^3} \right)\int_{\gamF} \theta_f\left(\nabla \theta_f \cdot \textbf{n}\right)\,d\gamF \, .
    \end{align*}
Choosing $\lambda_1 = \lambda_3 \left(\hat{d}/\epsilon_T^3 \right)$ allows both terms on the right-hand-sides  to vanish.
In summary, we choose $\lambda_2 = \hat{d}^3/\textrm{Da}$,  $\lambda_1 = \lambda$, and $\lambda_3 = \left(\epsilon_T^3/\hat{d}\right)\lambda$.
Importantly, there is now only a single free parameter $\lambda$.

With our choices for the coupling parameters and the functional energy
    \begin{align*}
    2\,E(t) = \frac{1}{\textrm{Pr}_f}\|\vufp\|^2_f+\frac{\hat{d}^3}{\chi\,\textrm{Pr}_m}\|\vump\|^2_m + \lambda\|\theta_f\|_f^2+\lambda \frac{\epsilon_T^3}{\hat{d}}\,\varrho\, \|\theta_m\|_m^2 \, ,
    \end{align*}
we are left with
    \begin{align} \label{eq:dedt}
    \frac{dE}{dt} = -\mathcal{D}+\mathcal{I} - \int_{\Gamma_i}\sum_{i=1}^{2}\frac{\hat{d}\,\alBJSJ}{\sqrt{\textrm{Da}}}\left(\vufp \cdot \boldsymbol{\tau}_i \right)^2\,d\gamF \leq -\mathcal{D} + \mathcal{I} \, ,
    \end{align}
where the definite and indefinite terms, $\mathcal{D}$ and $\mathcal{I}$, respectively, are defined as
    \begin{align*}
    \mathcal{D} &= \|\nabla\vufp\|^2_f + \frac{\hat{d}^3}{\textrm{Da}}\|\vump\|_m^2  + \lambda \| \nabla \theta_f\|_f^2 + \lambda\frac{\epsilon_T^3}{\hat{d}} \|\nabla \theta_m\|_m^2 \, ,\\
    \mathcal{I}&=-\left[\Raflam+\lambda \right]\left(w_f,\theta_f \right)_f -\left[\frac{\hat{d}^3}{\textrm{Da}}\Ramlam+\lambda \frac{\epsilon_T^3}{\hat{d}}\right]\left(w_m,\theta_m \right)_m \, ,
    \end{align*}
and $2\langle \mathbb{D}(\vufp), \mathbb{D}(\vufp)\rangle_f = \|\nabla \vufp\|_f^2$.
We are now using the notation $\Raflam$ and $\Ramlam$ to indicate dependence on the coupling parameter $\lambda$.
The change in the total energy of the system is bounded by
    \begin{align*}
    \frac{dE}{dt} \leq -\mathcal{D}+\mathcal{I} = \mathcal{D}\left(\dfrac{\mathcal{I}}{\mathcal{D}}-1\right) \leq \mathcal{D}\left(\max_{\mathcal{H}}\dfrac{\mathcal{I}}{\mathcal{D}}-1\right) = -\mathcal{D}\left(1-\max_{\mathcal{H}}\dfrac{\mathcal{I}}{\mathcal{D}}\right) \, ,
    \end{align*}
where $\mathcal{H}$ is the set of admissible solutions to equations $(\ref{nondimcoupFree})$ and $(\ref{nondimcoupPor})$ subject to $(\ref{nondimcoupInter})$.
Defining $\crn$ as the maximum of the ratio of energies
    \begin{align}
    \label{REdefn}
    \dfrac{1}{\crn} =\max_{\mathcal{H}}\dfrac{\mathcal{I}}{\mathcal{D}}
    \end{align} yields
    \begin{align}
    \frac{dE}{dt}\leq -\mathcal{D}\left(\dfrac{\crn - 1}{\crn} \right) \, .
    \end{align}
\setstcolor{red}
The Poincar\'e inequality implies that  $\mathcal{D}\geq cE$ for some constant $c>0$ \cite{hill2010b,hill2010a,hill2009}.
Then, if $\crn\ge 1$, Gronwall's inequality produces at least exponential convergence:
    \begin{align}
    \label{Ebound}
    E(t)\leq E(0) e^{-\hat{a}t}\rightarrow 0 \textrm{  as  }t\rightarrow \infty
    \end{align}
where $\hat{a} = c\left(\crn - 1 \right)/\crn$. Hence, the system is nonlinearly stable as long as $\crn\ge 1$.

$\crn = 1$ corresponds to the sharpest threshold for nonlinear stability that is made possible by \eqref{Ebound}, and hence is the most important case to analyze.
Setting $\crn = 1$ in \eqref{REdefn} produces an optimization problem, $\max_{\mathcal{H}}(\mathcal{I}/\mathcal{D}) = 1$, that can be solved by the Euler-Lagrange equations:
    \begin{align}\label{sys:ELeqFluid}
    z\in (0,1): \quad\begin{dcases}
    &2\,\nabla^2 \vufp - \left(\Raflam+\lambda\right)\theta_f\mathbf{k} = \DparD{ L_f}{\mathbf{x}}\, ,\\
    &2\lambda\nabla^2\theta_f  -\left(\Raflam+\lambda\right)w_f  =0\, ,
    \end{dcases}
    \end{align}
    \begin{align}\label{sys:ELeqPor}
    z\in (-1,0): \quad \begin{dcases}
    &2\dfrac{\hat{d}^3}{\textrm{Da}}\vump + \left(\dfrac{\hat{d}^3}{\textrm{Da}}\Ramlam+\lambda\frac{\epsilon_T^3}{\hat{d}}\right)\theta_m \mathbf{k}=\DparD{L_m}{\mathbf{x}} \, ,\\
    &2\lambda\frac{\epsilon_T^3}{\hat{d}}\nabla^2\theta_m -\left(\dfrac{\hat{d}^3}{\textrm{Da}}\Ramlam+\lambda\frac{\epsilon_T^3}{\hat{d}}\right)w_m =0\, ,
    \end{dcases}
    \end{align}
where $L_f, L_m$ are Lagrange multipliers for the fluid region and porous medium, respectively.
Taking the double curl of the first equations of \eqref{sys:ELeqFluid} and \eqref{sys:ELeqPor}  to remove the Lagrange multipliers and using the normal mode representations once again, we obtain the systems for the fluid layer and porous medium, respectively:
    \begin{align}\label{sys:solvFluid}
    z\in (0,1): \quad \begin{dcases}
    &2\,{\matt{\left(D_f^2 - a_f^2 \right)^2w_f}}+a_f^2 \left(\Raflam+\lambda\right)\theta_f = 0\, ,\\
    &2\lambda\left(D_f^2 - a_f^2 \right)\theta_f  -\left(\Raflam+\lambda\right)w_f =0\, ,
    \end{dcases}
    \end{align}
    \begin{align}\label{sys:solvPor}
    z\in (-1,0): \quad\begin{dcases}
    &2\dfrac{\hat{d}^3}{\textrm{Da}}\left(D_m^2 - a_m^2 \right)w_m -a_m^2 \left(\dfrac{\hat{d}^3}{\textrm{Da}}\Ramlam+\lambda\frac{\epsilon_T^3}{\hat{d}}\right)\theta_m =0\, ,\\
    &2\lambda\frac{\epsilon_T^3}{\hat{d}} \left(D_m^2 - a_m^2 \right)\theta_m -\left(\dfrac{\hat{d}^3}{\textrm{Da}} \Ramlam + \lambda\frac{\epsilon_T^3}{\hat{d}}\right)w_m =0\, .
    \end{dcases}
    \end{align}
For the interface and boundary conditions, we use the same equations as the linear case \eqref{eq:nonDlinIC}--\eqref{eq:nonDlinBC} with the exception of the Lions condition (replacing its linear counterpart) given by
    \begin{align*}
        -\mathbf{n}\cdot\mathbb{T}\left(\vufp, \pi_f\right)\mathbf{n}+\frac{1}{2}|\vufp|^2 = \hat{d}^2\,\pi_m\, .
    \end{align*}
Equations \eqref{sys:solvFluid}--\eqref{sys:solvPor} with the interface/boundary conditions as noted above constitute a generalized eigenvalue problem for $\Ramlam$ (recall that $\Ramlam$ and $\Raflam$ are related through (\ref{eq:relation})).
For given wavenumber $a_m$, we solve for $\Ramlam$ numerically, once again using the Chebyshev tau-QZ method.
We then maximize over $\lambda$ to obtain the sharpest threshold for nonlinear stability,
    \begin{align*}
    \Ramax = \max_{\lambda} \Ramlam \,.
    \end{align*}
Once $\Ramax$ is found for a range of wavenumbers, we can construct the marginal stability curve $\left(a_m, \Ramax \right)$, below which we are guaranteed {\em nonlinear stability}.

\section{Results and discussion} \label{sect:results}
In this section, we first present the marginal stability curves produced by the linear and nonlinear analysis.
Next, we show show that the relative difference between linear and nonlinear curves scales like $\textrm{Da}^1$ for small Darcy numbers, while the absolute difference scales like $\textrm{Da}^2$.
We find similar scalings for the differences between marginal stability curves produced by the BJSJ versus the BJJ or BJ interface conditions.
Lastly, we comment on resulting streamline patterns for convection cells occupying the entire domain or remaining solely in the fluid region, and remark on the effect of certain parameters on stability.

\subsection{Marginal Stability Results}
The marginal stability curves in Figure 2 show the $\Ramax$ values which mark the transition from stability to instability for each wavenumber $a_m$.
Below the linear marginal stability curves, we are guaranteed \textit{linear} stability, while we are assured unconditional stability below the nonlinear marginal stability curves.
In the area between the two curves, nonlinear effects could potentially destabilize the system even though the background state is linearly stable, i.e.~a subcritical instability.
However, Figure 2 shows that the linear and nonlinear curves follow each other closely, suggesting that the impact of these nonlinear terms is small, at least during the onset of convection.
We have explored an extensive range of parameters (not shown here) with similar findings. We therefore conclude that the linear theory accurately describes the onset of convection, and that the region of potential subcritical instabilities is very small.
Furthermore, this result implies that the linear stability thresholds, which are generally simpler to compute, actually approximate unconditional or global stability of the system to a high level of accuracy.
\matt{We remark that the selected energy function $E(t)$ may not necessarily be the optimal choice. It is conceivable that an improved choice of $E(t)$ may reduce the gap between the linear and nonlinear curves even further.}

\matt{
In comparing the linear stability system \eqref{eq:nonDlinTop}--\eqref{eq:nonDlinBC} and the nonlinear stability system \eqref{sys:solvFluid}--\eqref{sys:solvPor}, we note the nonlinear stability system loses explicit dependence on the Prandtl number. This loss of $\textrm{Pr}$-dependence has been observed in a variety of other convective problems \cite{nield2017}.
While the linear system does contain $\textrm{Pr}$ terms, they are all associated with the $\sigma$ eigenvalue terms. In the single layer case, the principle of exchange of stabilities implies that $\sigma$ is real as long as the Rayleigh number is positive, meaning again that there is no dependence on $\textrm{Pr}$. In the coupled case, however, exchange of stabilities has not been established rigorously. Our numerics indicate that $\sigma$ can indeed take complex values. However, we always observe $\sigma$ to be real whenever the Rayleigh number is positive. This numerical observation suggests that the principle of exchange of stabilities holds in practice, and consequently dependence on $\textrm{Pr}$ is lost.}

    \begin{figure}[!h]
    \centering
        \begin{subfigure}[b]{0.35\textwidth}
          \includegraphics[width=\textwidth]{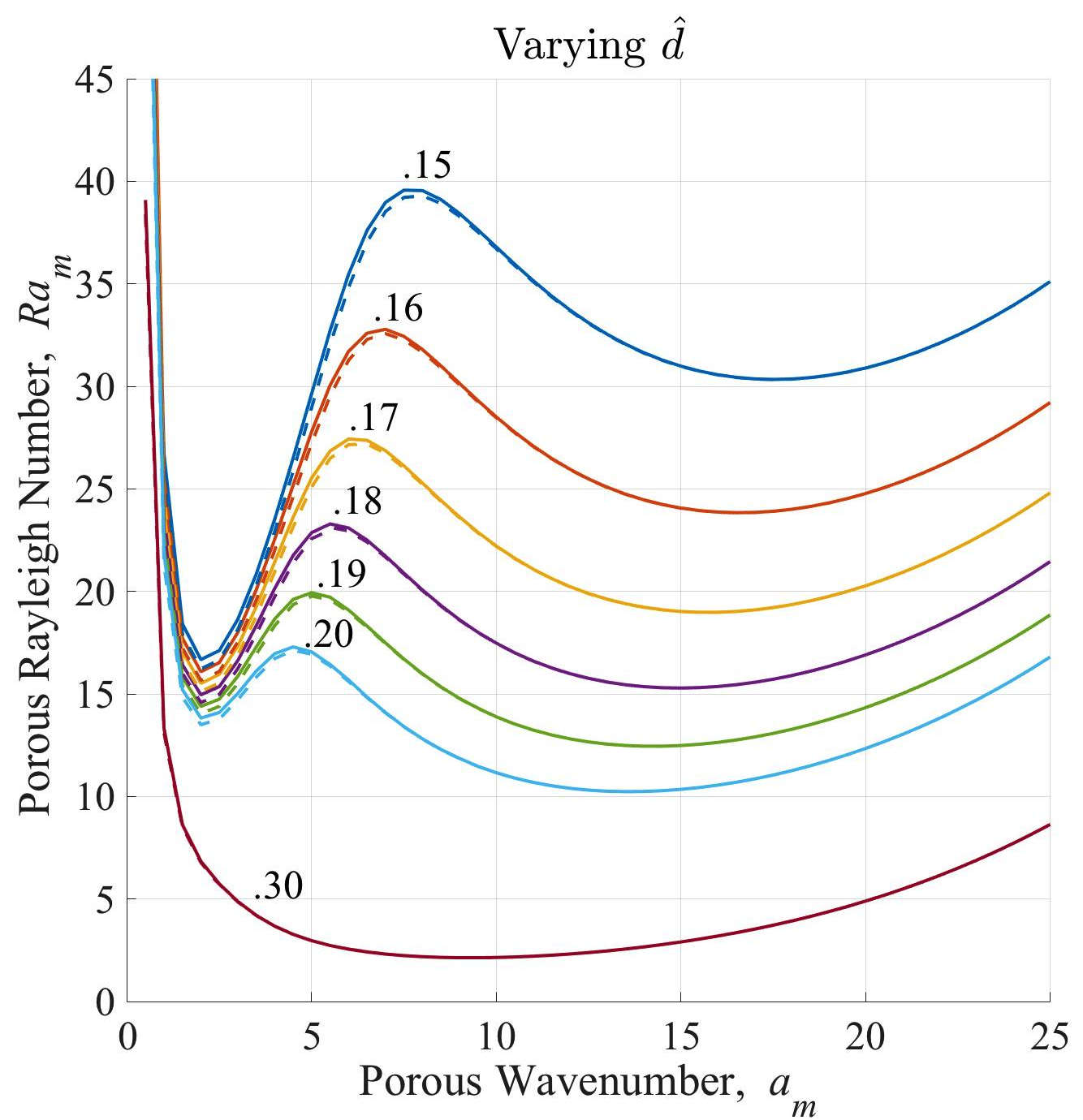}
          \caption{$\sqrt{\textrm{Da}} =5.0\times 10^{-3}$, $\alBJSJ=1.0$.\label{fig:compar1}}
        \end{subfigure}
        \hspace{.5in}
        \begin{subfigure}[b]{0.35\textwidth}
          \includegraphics[width=\textwidth]{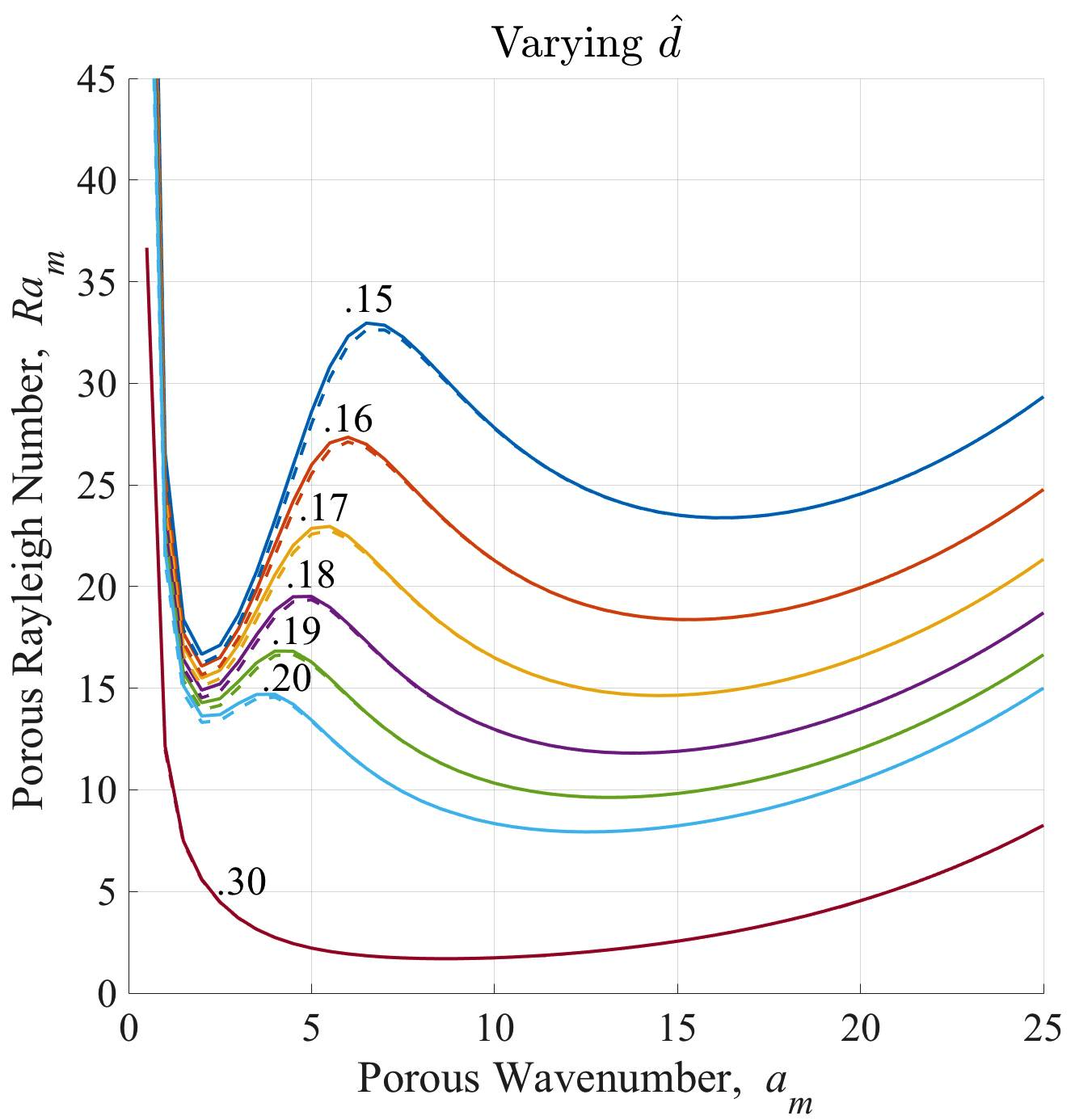}
          \caption{$\sqrt{\textrm{Da}} =5.0\times 10^{-3}$, $\alBJSJ=0.1$.\label{fig:compar2}}
        \end{subfigure}
        \begin{subfigure}[b]{0.35\textwidth}
          \includegraphics[width=\textwidth]{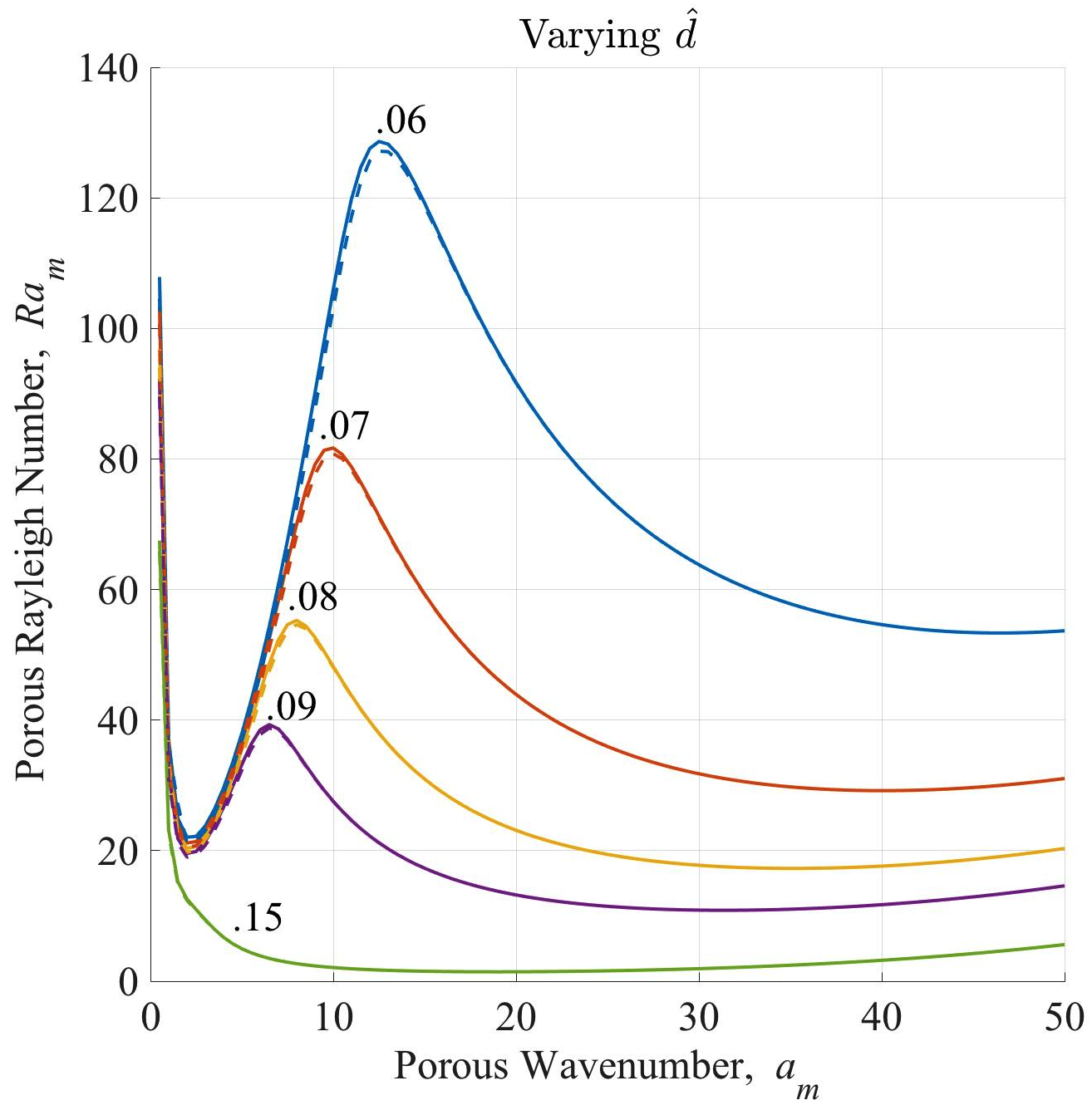}
          \caption{$\sqrt{\textrm{Da}} =1.0\times 10^{-3}$, $\alBJSJ=1.0$.\label{fig:compar3}}
        \end{subfigure}
        \hspace{.5in}
        \begin{subfigure}[b]{0.35\textwidth}
          \includegraphics[width=\textwidth]{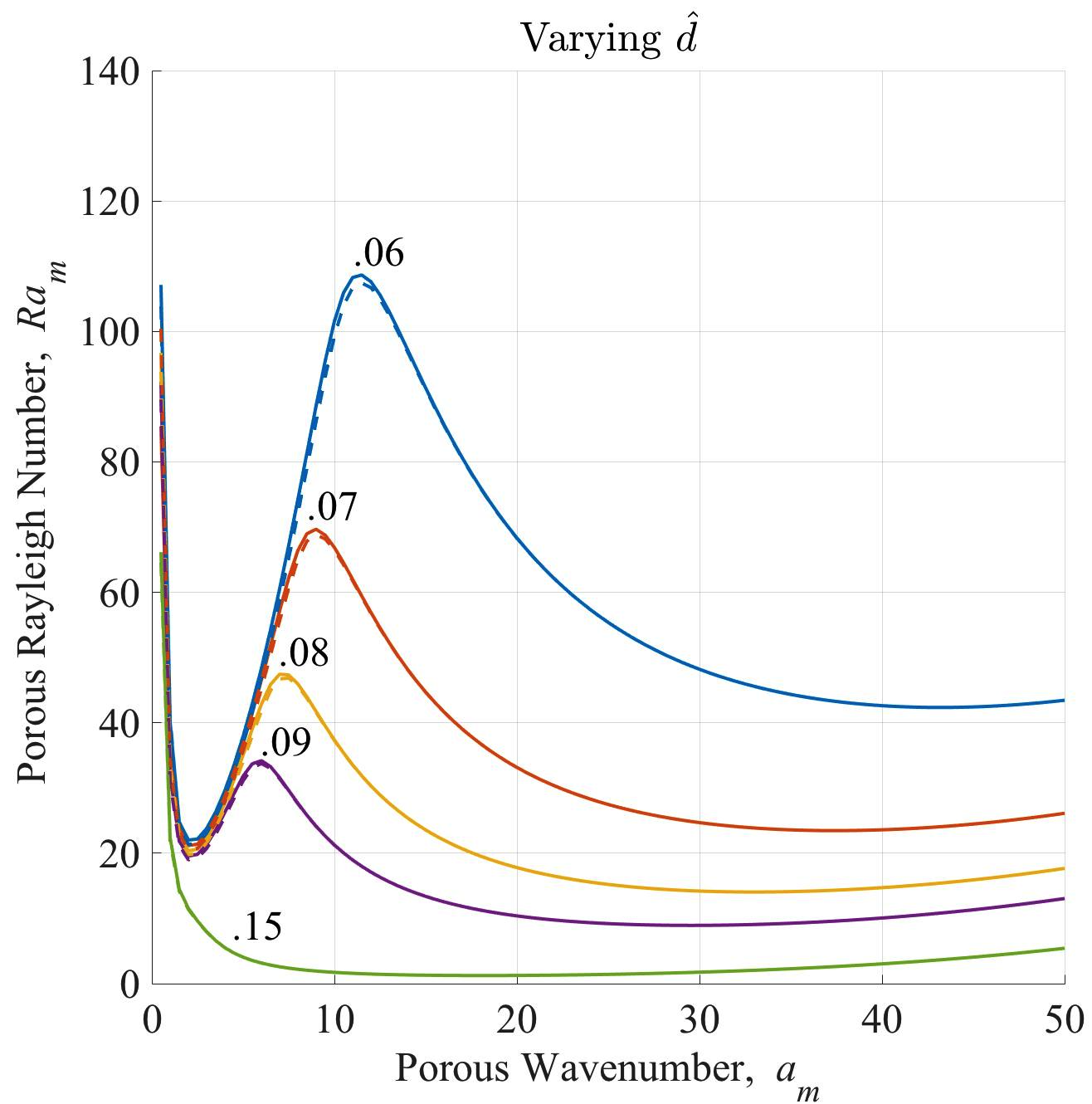}
          \caption{$\sqrt{\textrm{Da}} =1.0\times 10^{-3}$, $\alBJSJ=0.1$.\label{fig:compar4}}
        \end{subfigure}
        \caption{Marginal stability curves for different values of $\textrm{Da}$ and $\alBJSJ$, with $\epsilon_T=.7$ in all cases. Both linear (solid) and nonlinear (dashed) stability results are shown. The two results agree closely with one another in all cases, indicating that the region of potential subcritical instability is very small.
        \label{fig:compar}}
      \end{figure}

\subsection{Influence of Interface Conditions}
To quantify how closely the linear and nonlinear marginal stability thresholds agree, we examine the relative difference between the respective stability curves.
The Lions condition is used to produce the nonlinear thresholds while its linear counterpart is used as the normal interface condition for the linear stability curves.
The results are shown in Figure \ref{fig:nonVlin}.
\matt{For a fixed $\hat{d}$ and $a_m$, the $\Ram$ values are computed with the linear and nonlinear arguments for various $\textrm{Da}$ values. We then examine the relative difference between the two computed $\Ram$ values. }
For small Darcy numbers, $\textrm{Da}\in [10^{-8}, 10^{-4}],$ we see the relative difference scales like $\textrm{Da}^1$, as shown with the comparison line.
Given that $\Ram \sim \mathcal{O}\left(\textrm{Da}\right)$ in the small Darcy limit, the absolute difference scales like $\textrm{Da}^2$, as is expected from the theory and reflected in the numerical tests.
In the appendix, we present an asymptotic argument showing that, while the dynamic pressure term is $\mathcal{O}\left(\textrm{Da}\right)$, it only begins to affect the solutions at $\mathcal{O} \left(\textrm{Da}^2 \right)$.
Though somewhat heuristic, this asymptotic analysis provides guidance for the scaling of stability threshold differences found using the Lions and the linear interface conditions \matt{ in the small Darcy number regime.}

    \begin{figure}[!h]
    \centering
        \begin{subfigure}[b]{0.45\textwidth}
          \includegraphics[width=\textwidth]{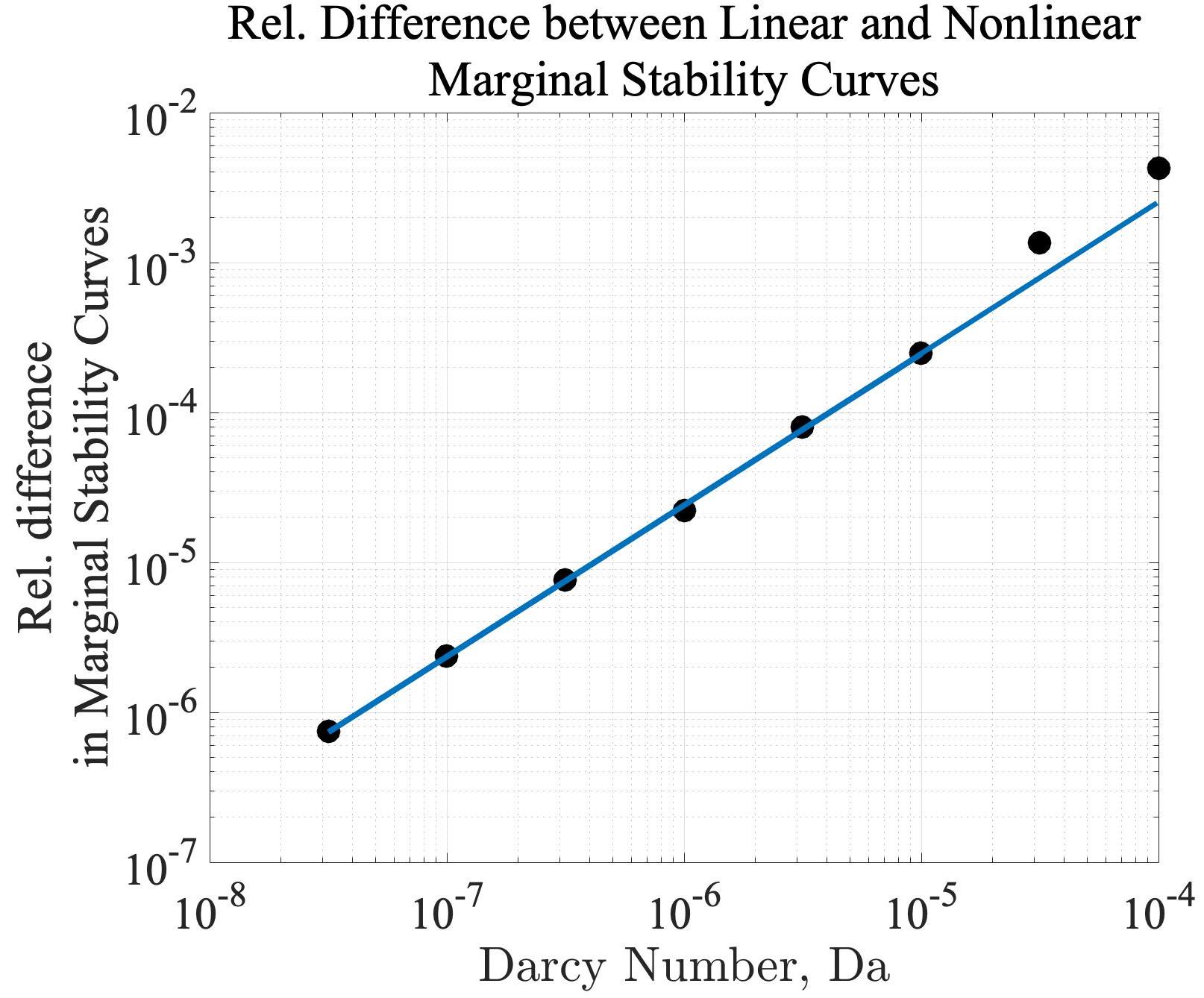}
          \caption{The relative difference between nonlinear and linear marginal stability curves produced with the Lions interface condition and its linear counterpart, respectively (with BJSJ used). \label{fig:nonVlin}}
        \end{subfigure}\quad
        \begin{subfigure}[b]{0.45\textwidth}
          \includegraphics[width=\textwidth]{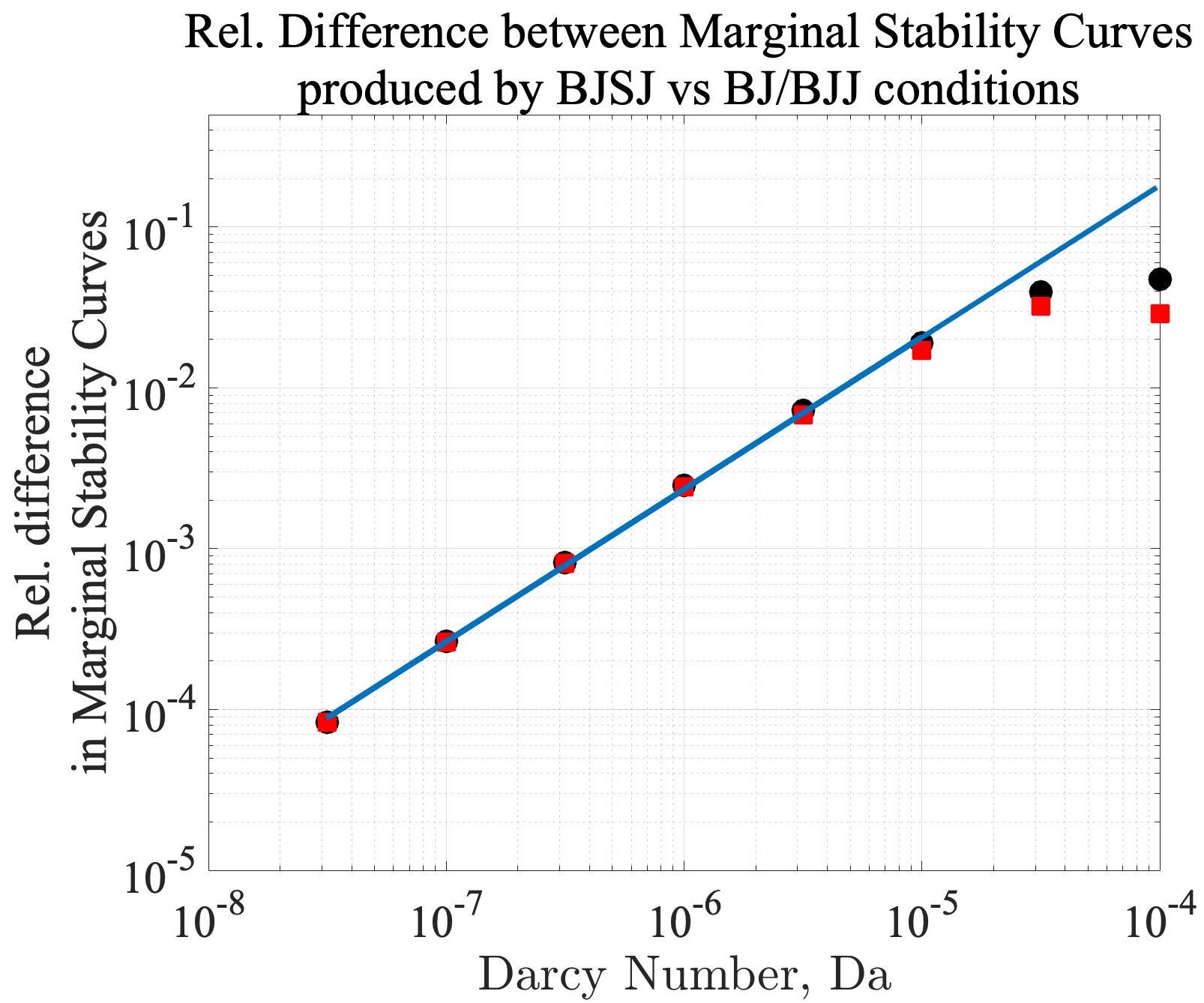}
          \caption{The relative differences between linear marginal stability curves produced with the: BJSJ and Jones interface conditions (black circles), and the BJSJ and Beavers-Joseph conditions (red squares).\label{fig:rel}}
        \end{subfigure}
        \caption{ Parameters: $\hat{d}=.1$, $a_m =25.0$, $\epsilon_T = .7$, $\alpha=1.0$. Both comparison lines have slope of $1$. \label{fig:err}}
    \end{figure}

Now, we briefly discuss small differences in the tangential interface conditions.
In particular, we show in Figure \ref{fig:rel}, the relative differences in the linear stability curves produced by the BJ, BJJ, and BJSJ interface conditions.
The relative differences between BJSJ and BJJ are marked with black circles while BJSJ versus BJ are marked with red squares.
Both of the relative differences scale like $\textrm{Da}^1$ and both absolute differences scale like $\textrm{Da}^2$.
Thus, using any of the three conditions results in similar qualitative behavior in the marginal stability curves.

An important parameter that enters these tangential interface conditions is the frictional coefficient $\alBJSJ$.
Looking back at Fig.~2, we vary $\alBJSJ$ from 1.0 to 0.1 in going from the left columns, (a) and (c), to the right, (b) and (d).
We note that although the marginal stability curves are altered, the location of their minima does not change significantly.
This minimum value of $\Ramax$ is known as the critical Raleigh number
    \begin{align}\label{eq:critRayNum}
    \Ramc = \min_{a_m^2} \Ramax \, ,
    \end{align}
which is the smallest Rayleigh number for which an unstable mode exists.
Thus, the critical Raleigh number exhibits low sensitivity to $\alBJSJ$, as is consistent with previous studies \cite{chen2010}.

\subsection{Fluid-dominated versus full convection}
An important insight that can be obtained from the marginal stability curves is whether the convection extends throughout the domain or is confined to the fluid region.
For this, we examine the wavenumber associated with $\Ramc$, which offers information on the lengthscale and aspect ratio of this most unstable mode; i.e.~smaller wavenumbers correspond to larger convection cells that extend throughout the domain while large wavenumbers correspond to smaller convection cells which arise only in the free-zone.
For example, in Figure \ref{fig:compar1}, for $\hat{d} = [.15, .18]$, we find the minima of the marginal stability curves all occur around $a_m = 2.0$. At $\hat{d} = .19$ though, the minimum shifts to a higher wavenumber, $a_m = 14.0$.
At some depth ratio between $\hat{d} = .18$ and $\hat{d}=.19$, the convection cells' aspect ratio suddenly changes from wide cells ($a_m=2.0$) to thin cells ($a_m=14.0$).
This phenomenon is also observed in \cite{straughan2002a}.
When the convection cells occupy both the porous medium and fluid region, we denote this as {\em full} convection while we use \textit{fluid-dominated} convection to describe when convection cells lie only in the fluid region.
Qualitatively, when the $\Ramc$ occurs at smaller wavenumber, we have full convection, and when $\Ramc$ occurs at larger wavenumber, we have fluid-dominated convection.

    \begin{figure}[!ht]
      \centering
    {\includegraphics[width=1.0\textwidth]{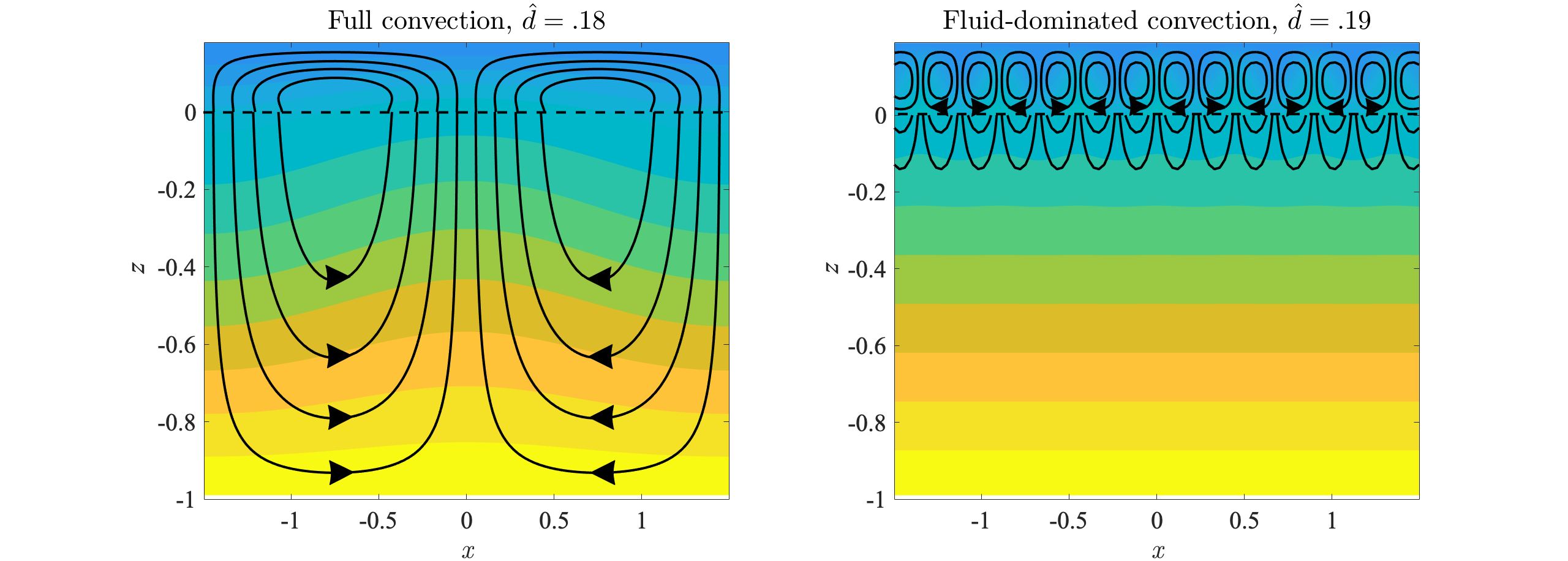}}
       \caption{Marginally stable flow configurations and temperature profiles (color) for two values of $\hat{d}$. (a) $\hat{d}=.18$ produces convection cells that extend throughout the entire domain, while (b) $\hat{d}=.19$ produces cells that are confined to the free-flow region. In both cases, $\sqrt{\textrm{Da}}=5.0\times 10^{-3}$, $\epsilon_T=.7$, $\alBJSJ=1.0$, and the BJSJ condition is used $\left(\Psi_J=1,\Psi_S = 0 \right)$.
       \label{fig:stream}}
    \end{figure}

    \begin{figure}[!ht]
      \centering
    {\includegraphics[width=0.95\textwidth]{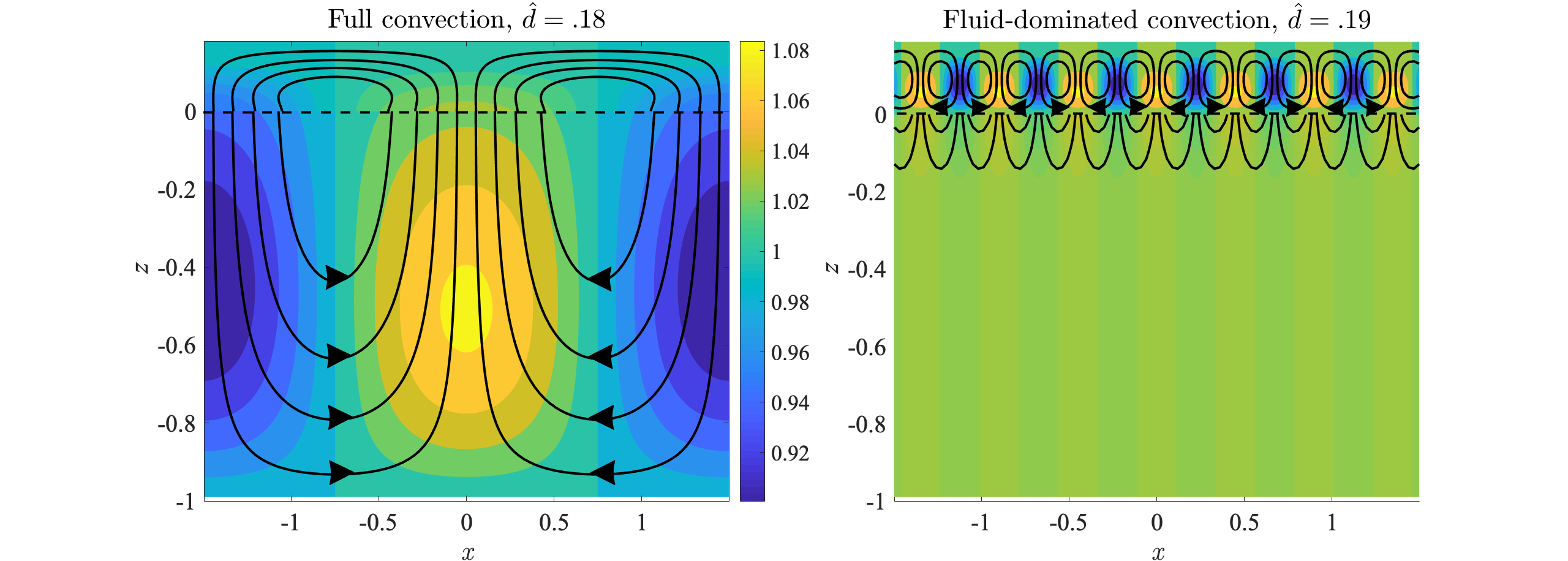}}
      \caption{Color map of the Nusselt number with streamlines in black for the same two cases shown in Fig.~\ref{fig:stream}. In the first case (a), the greatest variations of \textrm{Nu} occur in the porous medium, while in the second case (b) the extrema of $\textrm{Nu}$ are confined to the free-flow region.
      \label{fig:nuss}}
    \end{figure}

To understand which region dominates convection in a more quantitative sense, we examine the resulting streamline and temperature profiles, as well as the Nusselt numbers.
Figure \ref{fig:stream} shows the streamlines in black and the temperature profiles in color for $\hat{d}=.18$ and $\hat{d}=.19$ at their critical Rayleigh numbers with: $\chi=.3,$ $\sqrt{\textrm{Da}}= 5.0\times 10^{-3},$ $\alBJSJ=1.0$, $\epsilon_T =.7$.
\matt{The streamlines are computed via numerical solution of the linear system \eqref{eq:nonDlinTop}--\eqref{eq:nonDlinBC}.}
Both figures are plotted over the same $x$ range to more effectively show how the small change in the $\hat{d}$ value (from $\hat{d}=.18$ to $.19$) drastically alters the convection cells, streamlines, and temperature profiles.
For full convection, we see that the temperature and velocity deviations from the steady-state occur throughout the entirety of the domain.
\matt{For the fluid-dominated convection though, temperature and velocity fluctuations only occur in and immediately around the free zone.}
To further quantify these observations, we analyze the Nusselt number $\textrm{Nu}$, calculated from the vertical convective and conductive fluxes as
    \begin{align}\label{NusseltInfo}
        \textrm{Nu} = \frac{ J_{\textrm{cnv} } + J_{\textrm{cnd}}}{J_{\textrm{cnd}}}  \quad \textrm{ with }\quad   J_{\textrm{cnv}} = w_j\,T_j \quad \textrm{ and }\quad J_{\textrm{cnd}}=-\kappa_j\,\parD{T_j}{z_j}\,,
    \end{align}
for $j\in\{f,m\}$.
In Figure \ref{fig:nuss}, we show the Nusselt numbers for the same cases $\hat{d}=.18$ and $\hat{d}=.19$ with the same streamlines pictured in Figure \ref{fig:stream}.
In regions where $\textrm{Nu}=1,$ there is negligible vertical fluid flow and the heat transfer is purely conductive.
Wherever $\textrm{Nu}>1$, the convective flux is upward and it enhances the conductive flux.
On the other hand, when $\textrm{Nu}<1$ convective flux is downward which opposes the conductive flux.
At the middle of the convection cells and at the top and bottom of the domain, fluid motion is almost purely horizontal and so the Nusselt number is nearly $1$.
At the edges of the convection cells, we see $\textrm{Nu}$ attains its maximum and minimum as the flow is almost solely in the vertical direction, moving upward and downward for the maximum and minimum of $\textrm{Nu}$, respectively.
When the Nusselt number achieves its extrema in the fluid region, the convection is fluid-dominated while we have full convection when the Nusselt number varies throughout the whole domain.

With the analysis above, determining which region dominates convection is relatively straightforward.
However, determining parameter values where the convection shifts from full to fluid-dominated is more complicated.
The region that dominates convection depends on a number of parameters, namely the depth ratio $\Hat{d}$, the Darcy number $\textrm{Da}$, and the ratio of thermal diffusivities $\epsilon_T$.
For example, fixing $\textrm{Da}$ and $\epsilon_T$, one could compute the marginal stability curves for a number of $\Hat{d}$ values to find the depth ratio where the transition in convection occurs.
However, this can be a computationally demanding task, since, even producing a single marginal stability curve requires a search over the parameters $a_m$ and $\Ram$.
We therefore offer a simplified theory to determine whether the onset of convection is full or fluid-dominated.
Although the Darcy number $\textrm{Da}$ and the ratio of thermal diffusivities $\epsilon_T$ could also trigger the transition, we focus on the influence of $\Hat{d}$ in this paper.

\matt{For the purpose of developing a simplified theory, let us briefly consider the free-flow and porous domains as {\em uncoupled}.
As before, the Rayleigh number in each domain is denoted $\Ram$ and $\Raf$, with the same relationship as in \eqref{eq:RafRamrelation}:
    \begin{align} \label{eq:RafRamrel2}
     \Ram = \Raf \frac{\textrm{Da}\,\epsilon_T^2}{\Hat{d}^4}
    \end{align}
Let $\Ramcu$ and $\Rafcu$ denote the corresponding critical values in the {\em uncoupled} system. Both of these values are well known, $\Rafcu=1707$ and $\Ramcu=4\pi^2$.
Heuristically, the strength of convection in each domain should be proportional to the Rayleigh number scaled by the appropriate critical value.
Thus, the transition from full to fluid-dominated convection is expected to occur approximately where the ratios are equal,}
    \begin{align*}
     {\matt{\frac{\Ram}{\Ramcu} = \frac{\Raf}{\Rafcu}\, .}}
    \end{align*}
\matt{Substituting relationship \eqref{eq:RafRamrel2} and solving for $\hat{d}$ gives the predicted transition value}
    \begin{align}\label{eq:critical}
        {\matt{ \dhatc = \left[\frac{\Rafcu}{\Ramcu}\,\textrm{Da}\,\epsilon_T^2 \right]^{1/4}\,.}}
    \end{align}
Therefore, by neglecting any coupling between the two regions, we have obtained a simple, approximate formula for the depth ratio at which convection is predicted to transition from full to fluid dominated.

We test this theory with the parameters $\sqrt{\textrm{Da}} = 5.0\times 10^{-3}$ and $\epsilon_T = 0.7$.
Numerically, the transition occurs around $\hat{d}^*\approx .181$, as shown in Figure \ref{fig:compar1}. The value predicted by the simplified theory is
    \begin{align*}
        \left(\frac{1707}{4\pi^2}\,(5.0\times 10^{-3})^2\,(.7)^2\right)^{1/4} \approx .151 \, .
    \end{align*}
which agrees with the numerically computed value to within 16\% error.
Secondly, we test $\sqrt{\textrm{Da}} = 1.0 \times 10^{-3}$ and $\epsilon_T = 0.7$, with the result shown in Figure \ref{fig:compar3}.
The simple theory predicts the transition to occur at $\hat{d}^* \approx .067$, while numerics show the transition to occur around $\hat{d}^*\approx .079$, corresponding to an error of 15\%.

    \begin{table}[!h]\centering
    \begin{tabular}{|c|c|c|c|c|c|}
    \hline
    $\sqrt{\textrm{Da}}$ & $\epsilon_T$ & Predicted $\hat{d}^{*}$ & Actual $\hat{d}^{*}$ & Relative error& Figure reference \\ \hline
    $5.0\times 10^{-3}$ & 0.7 & .151 & .181 & 16.5\% & \ref{fig:compar1} \\ \hline
    $1.0\times 10^{-3}$ & 0.7 & .067 & .079 & 15.1 \%& \ref{fig:compar3} \\ \hline
    $5.0\times 10^{-3}$ & 0.5 & .128 & .155 & 17.4 \% & \ref{fig:et_0_5}\\ \hline
    $5.0\times 10^{-3}$ & 1.5 & .222 & .256 & 13.3 \% & \ref{fig:et_1_5} \\ \hline
    \end{tabular}\caption{Table of predicted and actual $\hat{d}^*$ values with relative errors. Fixed parameters: $\chi=.3,$ $\alBJSJ=1.0$.}
    \end{table}

Table 1 compares the predicted and actual values of $\Hat{d}$ for the cases discussed above as well as a few additional cases.
For various $\textrm{Da}$ and $\epsilon_T$ values, we see our theory, numerics, and intuition are all in agreement.
For example, with the last two rows of the table, as the ratio of thermal diffusivities $\epsilon_T = \lambda_f/\lambda_m$ decreases, convection occurs more easily through the entire domain.
Consequently, the transition from \matt{full} to fluid-dominated convection takes place at a lower depth ratio.
Figure \ref{fig:et_var} illustrates this trend with marginal stability curves for $\epsilon_T$ values of $0.5$ and $1.5$.

\matt{In summary, we find the transition depth $\dhatc$ heuristically predicted by \eqref{eq:critical} agrees with the true transition value to within about $15 \%$ in all cases tested. Thus, while the theory is not extremely accurate, it is a useful first estimate to narrow the parameter range that must be searched to find the transition depth. The theory is perhaps even more accurate than could be expected given that it completely neglects coupling between the two regions. It is a promising first step towards developing a more refined theory to predict the transition, perhaps by accounting for weak coupling between the two regions. We also remark that the simplified theory seems to consistently underpredict the transition depth. Thus, the effect of coupling is to inhibit fluid-dominated convection in favor of full convection.}

      \begin{figure}[!h]
    \centering
        \begin{subfigure}[b]{0.35\textwidth}
          \includegraphics[width=\textwidth]{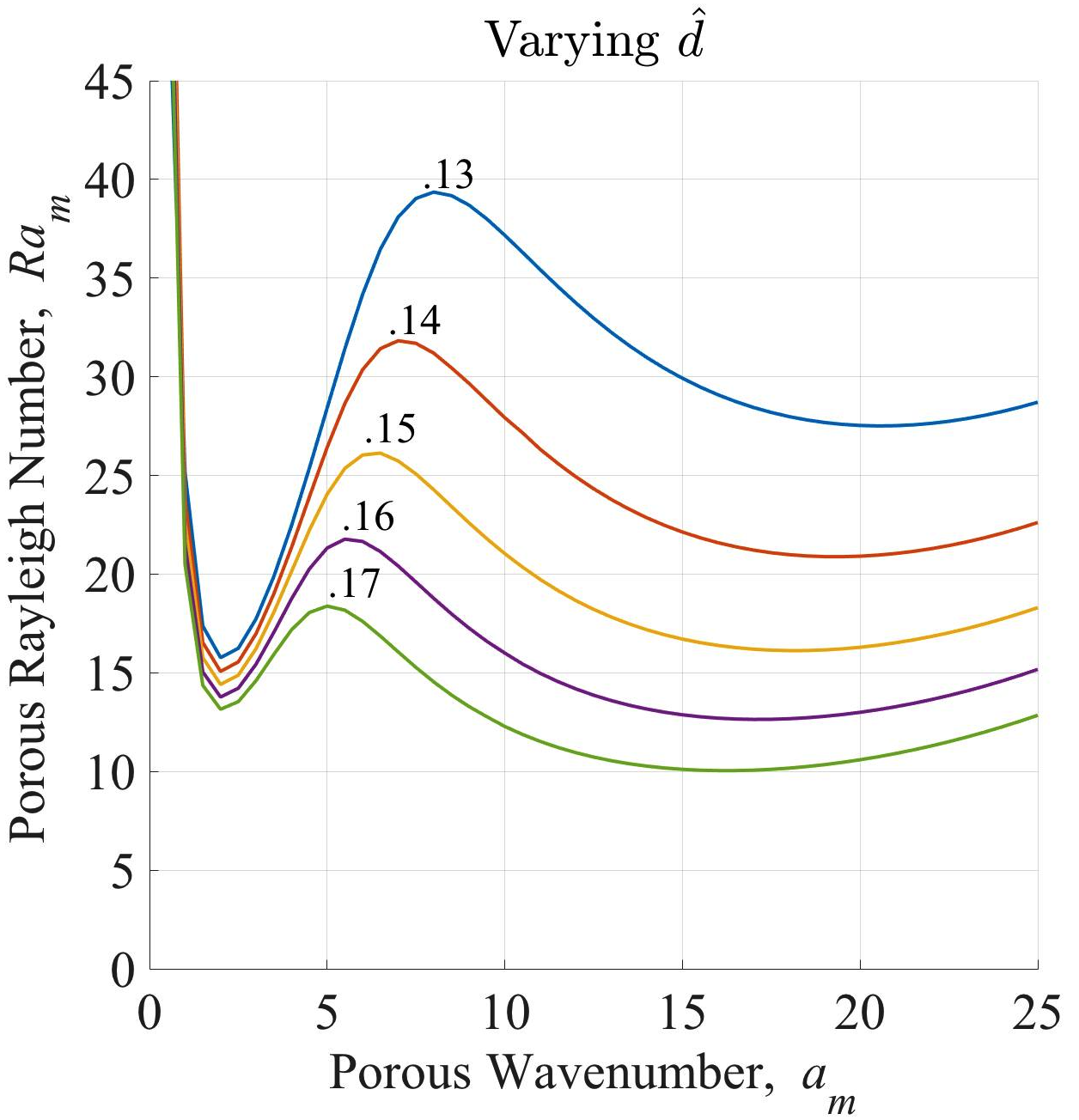}
          \caption{$\epsilon_T=0.5$. \label{fig:et_0_5}}
        \end{subfigure}
        \hspace{.5in}
        \begin{subfigure}[b]{0.35\textwidth}
          \includegraphics[width=\textwidth]{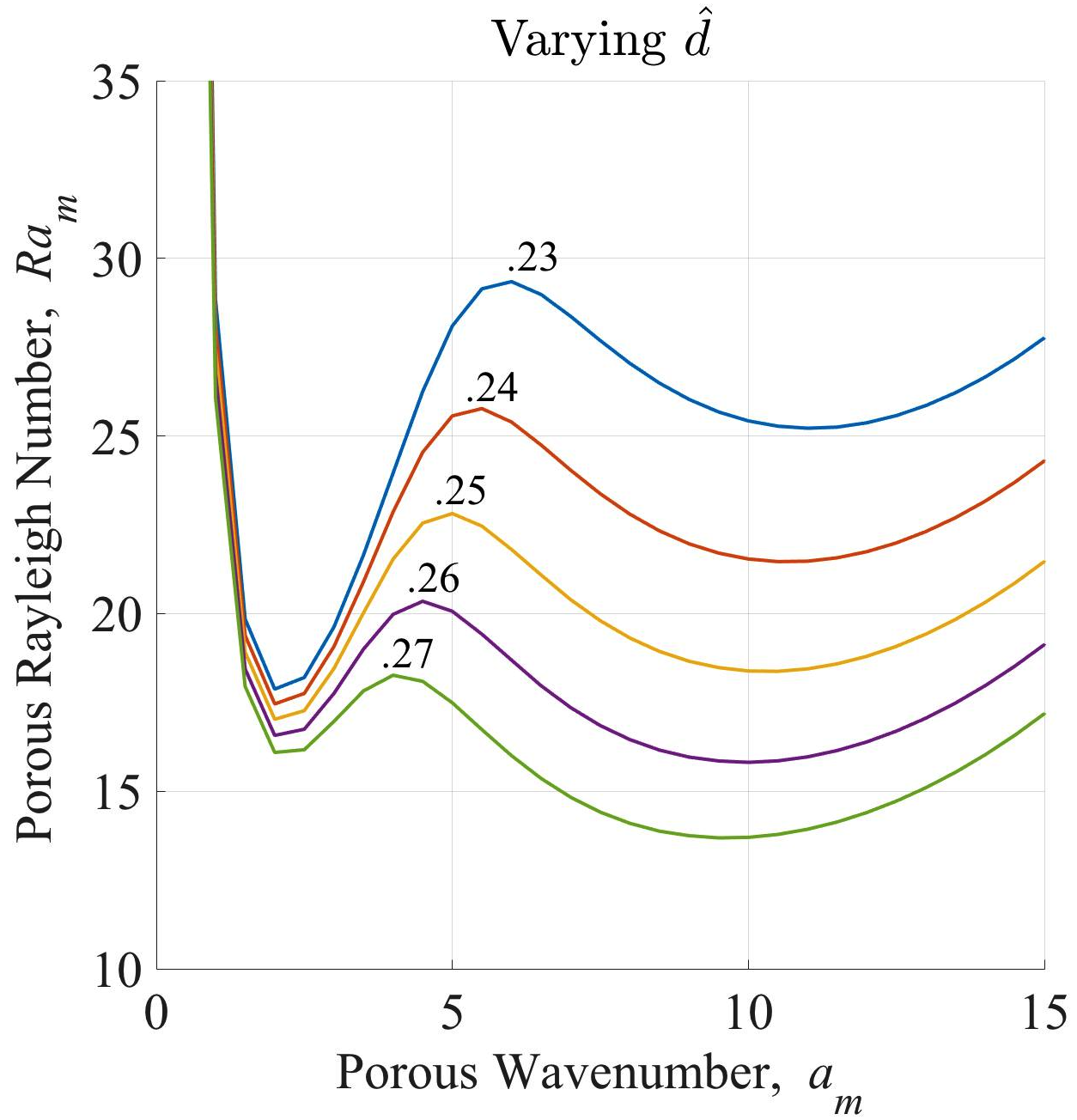}
          \caption{$\epsilon_T=1.5$. \label{fig:et_1_5}}
        \end{subfigure}
        \caption{Marginal stability curves for varying $\epsilon_T$, with $\sqrt{\textrm{Da}} =5.0\times 10^{-3}$ and $\alBJSJ=1.0$ fixed.
        Linear stability results with the BJSJ interface condition used $\left(\Psi_J=1,\Psi_S = 0 \right)$.\label{fig:et_var}}
    \end{figure}

\section{Conclusions}
In this work, we presented linear and nonlinear stability results of the coupled Navier-Stokes-Darcy-Boussinesq system that governs convection in a fluid-porous medium system. The main contribution is the newly obtained nonlinear analysis, which relies crucially on the Lions interface condition in order to establish an energy law.
We found that the marginal stability curves produced by the nonlinear and linear analysis follow each other closely, suggesting that linear stability is sufficient to describe the onset of convection. The agreement between the linear and nonlinear curves also implies that the more easily obtained linear thresholds indicate unconditional or global stability of the coupled fluid-porous system, at least for relatively small Darcy number.
Some additional results concerning convection are related to choosing interface conditions, namely those specifying tangential stress.
We showed the three different choices (BJ, BJJ, and BJSJ) are essentially the same, at least in terms of the onset of convection at small Darcy number regime; hence, it makes sense to adopt BJSJ due to the associated mathematical convenience.

We also postulated a simple theory to predict the transition from full to fluid-dominated convection due to changes in the depth ratio, the Darcy number, and the ratio of thermal diffusivities.
\matt{We find estimated transition depths to agree with the numerically computed values with reasonable accuracy (roughly $15\%$ error).
Accurate prediction of this transition could have applications in geophysics and in alloy solidification \cite{lebars2006a,lebars2006b}, and further refinement of the theory is an exciting future direction.}
In addition, while this work considered a flat, stationary interface between the free-zone and porous medium, future work could consider more complex interfaces \cite{allen1984, han2016decoupled}, or boundaries that move or evolve due to natural processes \cite{MooreCPAM2017, Quaife2018, zhang2000periodic}.

\section*{Acknowledgements}
M.N.J. Moore acknowledges the support of Simons Foundation Collaboration Grants for Mathematicians 524259 and X. Wang acknowledges the support of NSFC 11871159 and Guangdong Provincial Key Laboratory for Computational Science and Material Design 2019B030301001..

\bibliographystyle{siamplain.bst}
\bibliography{nlsa_paper}

\nomenclature[A,01]{$f$}{Fluid region}
\nomenclature[A,02]{$m$}{Porous medium}
\nomenclature[A,03]{$s$}{Solid component of the medium}

\nomenclature[B, 01]{$\rho_0$, $\mu_0$, $\nu$}{Reference density, dynamic viscosity, and kinematic viscosity of the fluid}
\nomenclature[B, 02]{$\beta$}{Coefficient of thermal expansion}
\nomenclature[B, 03]{$g$}{Gravitational constant }
\nomenclature[B, 04]{$\Pi$, $\chi$}{Permeability and porosity of the medium}
\nomenclature[B, 05]{$d_j$}{Depth of the $j\in\{f,m\}$ region}
\nomenclature[B, 05]{$T_U$, $T_L$, $T_0$}{Temperature at the top of the domain, bottom of the domain, and at the interface}
\nomenclature[B, 07]{$\kappa_j$}{Thermal conductivity of the $j\in\{f,m,s\}$ region }
\nomenclature[B,08]{$\left(c_p\rho_0\right)_j$}{Specific heat capacity of the $j\in\{f,m,s\}$ region}
\nomenclature[B,09]{$\lambda_j$}{Thermal diffusivity of the $j\in\{f,m\}$ region, $\lambda_j=\kappa_j/\left(c_p\rho_0\right)_f$}

\nomenclature[C, 01]{$\textbf{n}$, $\boldsymbol{\tau}$, $\textbf{k}$  }{Unit normal, unit tangent, and upward pointing unit normal }
\nomenclature[C, 02]{$\textrm{Da}$}{Darcy number, $\textrm{Da}=\Pi/d_m^2$}
\nomenclature[C, 03]{$\alpha$}{Empirically determined coefficient in BJSJ condition}
\nomenclature[C, 03]{$\Psi_J, \Psi_S$}{Switches associated with the BJSJ and related interface conditions }
\nomenclature[C, 04]{$\hat{d}$}{Ratio of depths, $\hat{d}=d_f/d_m$}
\nomenclature[C, 05]{$\epsilon_T$}{Ratio of thermal diffusivities, $\epsilon_T=\lambda_f/\lambda_m$}
\nomenclature[C, 06]{$\varrho$}{Inverse of the ratio of specific heat capacities, $\varrho=\left(c_p\rho_0\right)_m/\left(c_p\rho_0\right)_f$}
\nomenclature[C, 07]{$\textrm{Pr}_j$}{Prandtl number of $j\in\{f,m\}$ region, $\textrm{Pr}_j=\nu/\lambda_j$}
\nomenclature[C, 11]{$\lambda, \lambda_1, \lambda_2, \lambda_3$}{Coupling parameters used in nonlinear stability analysis }
\nomenclature[C, 12]{$\crn$}{Maximum of the ratio of $\mathcal{I}$ and $\mathcal{D}$ energies, defined in \eqref{REdefn} }
\nomenclature[C, 12]{$\hat{a}$}{Coefficient $\hat{a} = c\left(\crn - 1 \right)/\crn$, used in  \eqref{Ebound}}
\nomenclature[C, 14]{$\hat{d}^*$}{Critical depth ratio marking transition from full to fluid-dominated convection}

\nomenclature[D, 01]{$F$}{Horizontal planform (or planeform tiling), $F=F(x,y)$ }
\nomenclature[D, 02]{$a_j$}{Horizontal wavenumber of $j\in\{f,m\}$ region }
\nomenclature[D, 03]{$\sigma_j$}{Eigenvalue of $j\in\{f,m\}$ region in linear stability analysis }
\nomenclature[D, 06]{$E(t)$}{Functional energy in the nonlinear stability analysis }
\nomenclature[D, 07]{$\mathcal{I}$, $\mathcal{D}$}{Indefinite and definite terms in nonlinear stability analysis }
\nomenclature[D, 09]{$\textrm{Nu}$}{Nusselt number, defined in \eqref{NusseltInfo}}
\nomenclature[D, 10]{$J_{\textrm{cnv}},J_{\textrm{cnd}} $}{Vertical convective and conductive fluxes, defined in \eqref{NusseltInfo}  }

\nomenclature[E, 03]{$\textrm{Ra}_j$}{Rayleigh number of $j\in\{f,m\}$ region, defined in \eqref{eq:RafRamrelation}}
\nomenclature[E, 04]{$\Raflam, \Ramlam$}{Rayleigh numbers with dependence on coupling parameter $\lambda$ in nonlinear stability analysis }
\nomenclature[E, 05]{$\Ramc$}{Critical Rayleigh number porous medium, defined in \eqref{eq:critRayNum}}
\nomenclature[E, 13]{$\Ramcu, \Rafcu$}{Critical Rayleigh numbers of the {\em uncoupled} regions}

\nomenclature[F, 01]{$\Omega$}{Total domain}
\nomenclature[F, 02]{$\Omega_j$}{$j\in\{f,m\}$ region of the domain}
\nomenclature[F, 03]{$\Gamma_j$}{Interface, from the $j\in\{f,m\}$ region }
\nomenclature[F, 04]{$D_j$}{Derivative with respect to $z_j$, $j\in\{f,m\}$ region }
\nomenclature[F, 05]{$\nabla_H^2$}{Horizontal Laplacian operator}

\newpage
\begin{mdframed}
\printnomenclature[2cm]
\end{mdframed}

\newpage
\section*{Appendix II: Asymptotics}
The dynamic pressure term of the Lions interface condition specifying the balance of force in the normal direction is small.
As a result, the difference between solutions produced with the Lions interface condition and its linear counterpart is heuristically small as well.
However, this claim has been speculative until now.
With a formal asymptotic argument, we show that the size of the dynamic pressure term is $\mathcal{O}\left(\textrm{Da}\right)$ in the limit as the small Darcy number limit.
Additionally, we find that this term begins to affect solutions to the perturbed systems at $\mathcal{O}\left(\textrm{Da}^2 \right)$

With $\textrm{Da} = \varepsilon^2\rightarrow 0$ (and all other constants held constant), we employ the ansatz that our solutions take the form:
\begin{align*}
\vup_j^{\varepsilon} = \vup_j^{(0)} + \varepsilon\, \vup_j^{(1)} + \varepsilon^2 \,\vup_j^{(2)} +\hdots,\\
\pi_j^{\varepsilon} = \pi_j^{(0)} + \varepsilon\, \pi_j^{(1)} + \varepsilon^2 \,\pi_j^{(2)} +\hdots,\\
\theta_j^{\varepsilon} = \theta_j^{(0)} + \varepsilon \,\theta_j^{(1)} + \varepsilon^2\, \theta_j^{(2)} +\hdots,
\end{align*} for $j\in \{f,m\}$. We have the components of $\vup_\textbf{f}^{\varepsilon}=(u_f^{\varepsilon},v_f^{\varepsilon},w_f^{\varepsilon})$ where
\begin{align*}
u_f^{\varepsilon} = u_f^{(0)} + \varepsilon \,u_f^{(1)} + \varepsilon^2 \,u_f^{(2)} + \hdots,
\end{align*} with the components of $\vup_\textbf{m}^{\varepsilon}$ defined in the same fashion. Substituting our ansatz into systems (\ref{nondimcoupFree}), (\ref{nondimcoupPor}), (\ref{nondimcoupInter}):

In $\Omega_f$:
\begin{equation*}
\begin{dcases}
&\frac{1}{\textrm{Pr}_f}\parD{}{t}\left[\vufp^{(0)} + \varepsilon\, \vufp^{(1)} + \varepsilon^2 \,\vufp^{(2)} +\hdots \right] +\left[\vufp^{(0)} + \varepsilon\, \vufp^{(1)} + \varepsilon^2\, \vufp^{(2)} +\hdots \right]\cdot \nabla \left[\vufp^{(0)} + \varepsilon\, \vufp^{(1)} + \varepsilon^2\, \vufp^{(2)} +\hdots \right] \\
&\hspace{1in}= \nabla^2 \left[\vufp^{(0)} + \varepsilon \,\vufp^{(1)} + \varepsilon^2\, \vufp^{(2)} +\hdots \right]-\nabla\left[ \pi_f^{(0)} + \varepsilon\, \pi_f^{(1)} + \varepsilon^2\, \pi_f^{(2)} +\hdots\right]\\
&\hspace{1.15in}- \Raf\,\left[\theta_f^{(0)} + \varepsilon \,\theta_f^{(1)} + \varepsilon^2\, \theta_f^{(2)} +\hdots\right] \mathbf{k},\\
&\nabla \cdot \left[\vufp^{(0)} + \varepsilon\, \vufp^{(1)} + \varepsilon^2\, \vufp^{(2)} +\hdots \right] = 0,\\
&\parD{}{t}\left[\theta_f^{(0)} + \varepsilon \,\theta_f^{(1)} + \varepsilon^2\, \theta_f^{(2)} +\hdots\right] + \textrm{Pr}_f\,\left[\vufp^{(0)} + \varepsilon\, \vufp^{(1)} + \varepsilon^2\, \vufp^{(2)} +\hdots \right] \cdot \nabla \left[\theta_f^{(0)} + \varepsilon\, \theta_f^{(1)} + \varepsilon^2 \,\theta_f^{(2)} +\hdots\right]  \\
&\hspace{1in}= \nabla^2 \left[\theta_f^{(0)} + \varepsilon\, \theta_f^{(1)} + \varepsilon^2\, \theta_f^{(2)} +\hdots\right]  - \left[w_f^{(0)} + \varepsilon \,w_f^{(1)} + \varepsilon^2 \,w_f^{(2)} + \hdots \right].
\end{dcases}
\end{equation*}

In $\Omega_m$:
\begin{equation*}
\begin{dcases}
&\frac{1}{\chi}\frac{\varepsilon^2}{\textrm{Pr}_m}\, \parD{}{t}\left[\vump^{(0)} + \varepsilon\, \vump^{(1)} + \varepsilon^2 \,\vump^{(2)} +\hdots \right] + \left[\vump^{(0)} + \varepsilon\, \vump^{(1)} + \varepsilon^2 \,\vump^{(2)} +\hdots \right]\\
&\hspace{1in}= -\frac{\varepsilon^2}{\hat{d}^2}\,\nabla \left[\pi_m^{(0)} + \varepsilon\, \pi_m^{(1)} + \varepsilon^2 \,\pi_m^{(2)} +\hdots \right] - \varepsilon^2 \, \Raf \frac{\epsilon_T^2}{\hat{d}^4}\, \left[\theta_m^{(0)} + \varepsilon \,\theta_m^{(1)} + \varepsilon^2\, \theta_m^{(2)} +\hdots\right] \mathbf{k},\\
&\nabla \cdot \left[\vump^{(0)} + \varepsilon\, \vump^{(1)} + \varepsilon^2 \,\vump^{(2)} +\hdots \right] = 0,\\
&\varrho\, \parD{}{t}\left[\theta_m^{(0)} + \varepsilon \,\theta_m^{(1)} + \varepsilon^2\, \theta_m^{(2)} +\hdots\right] +\textrm{Pr}_m\,\left[\vump^{(0)} + \varepsilon\, \vump^{(1)} + \varepsilon^2 \,\vump^{(2)} +\hdots \right]\cdot \nabla\left[\theta_m^{(0)} + \varepsilon \,\theta_m^{(1)} + \varepsilon^2\, \theta_m^{(2)} +\hdots\right]   \\
&\hspace{1in}=\nabla^2 \left[\theta_m^{(0)} + \varepsilon \,\theta_m^{(1)} + \varepsilon^2\, \theta_m^{(2)} +\hdots\right]  -\left[w_m^{(0)} + \varepsilon \,w_m^{(1)} + \varepsilon^2 \,w_m^{(2)} + \hdots \right].
\end{dcases}
\end{equation*}

On $\Gamma_i$:
\begin{equation*}
\begin{dcases}
&\hat{d}\left[\theta_f^{(0)} + \varepsilon\, \theta_f^{(1)} + \varepsilon^2\, \theta_f^{(2)} +\hdots\right] = \epsilon_T^2\,\left[\theta_m^{(0)} + \varepsilon\, \theta_m^{(1)} + \varepsilon^2\, \theta_m^{(2)} +\hdots\right],\\
& \nabla \left[\theta_f^{(0)} + \varepsilon\, \theta_f^{(1)} + \varepsilon^2\, \theta_f^{(2)} +\hdots\right] \cdot \mathbf{n}= \epsilon_T \,\nabla \left[\theta_m^{(0)} + \varepsilon\, \theta_m^{(1)} + \varepsilon^2\, \theta_m^{(2)} +\hdots\right]\cdot \mathbf{n},\\
&\left[\vufp^{(0)} + \varepsilon\, \vufp^{(1)} + \varepsilon^2 \,\vufp^{(2)} +\hdots \right]\cdot \mathbf{n} = \left[\vump^{(0)} + \varepsilon\, \vump^{(1)} + \varepsilon^2 \,\vump^{(2)} +\hdots \right]\cdot \mathbf{n}\, , \\
& \varepsilon \,\boldsymbol{\tau} \cdot \mathbb{T}\left(\vufp^{(0)} + \varepsilon\, \vufp^{(1)} + \varepsilon^2 \,\vufp^{(2)} +\hdots, \pi_m^{(0)} + \varepsilon\, \pi_m^{(1)} + \varepsilon^2 \,\pi_m^{(2)} +\hdots \right)\mathbf{n}\\
&\hspace{1in}=\alBJSJ\,\left( \boldsymbol{\tau} \cdot \left[\vufp^{(0)} + \varepsilon\, \vufp^{(1)} + \varepsilon^2 \,\vufp^{(2)} +\hdots \right] \right)\textrm{ for }\gamma=1,2,\\
&-\mathbf{n}\cdot\mathbb{T}\left(\vufp^{(0)} + \varepsilon\, \vufp^{(1)} + \varepsilon^2 \,\vufp^{(2)} +\hdots,  \pi_f^{(0)} + \varepsilon\, \pi_f^{(1)} + \varepsilon^2\, \pi_f^{(2)} +\hdots \right)\mathbf{n}= \hat{d}^2\,\left[ \pi_m^{(0)} + \varepsilon\, \pi_m^{(1)} + \varepsilon^2\, \pi_m^{(2)} +\hdots\right],\\
&\textrm{or }-\mathbf{n}\cdot\mathbb{T}\left(\vufp^{(0)} + \varepsilon\, \vufp^{(1)} + \varepsilon^2 \,\vufp^{(2)} +\hdots,  \pi_f^{(0)} + \varepsilon\, \pi_f^{(1)} + \varepsilon^2\, \pi_f^{(2)} +\hdots \right)\mathbf{n}+\frac{1}{2}\left|\vufp^{(0)} + \varepsilon\, \vufp^{(1)} + \varepsilon^2 \,\vufp^{(2)} +\hdots \right|^2 \\
&\hspace{1in}= \hat{d}^2\,\left[ \pi_m^{(0)} + \varepsilon\, \pi_m^{(1)} + \varepsilon^2\, \pi_m^{(2)} +\hdots\right].
\end{dcases}
\end{equation*}

\subsection*{Balancing $\mathcal{O}(1)$}
\begin{equation*}
\textrm{ In }\Omega_f: \hspace{.15in}
\begin{dcases}
&\frac{1}{\textrm{Pr}_f}\parD{\vufp^{(0)}}{t}+\vufp^{(0)} \cdot \nabla\vufp^{(0)}= \nabla^2\vufp^{(0)} -\nabla\pi_f^{(0)}- \Raf\,\theta_f^{(0)}\,\mathbf{k},\\
&\nabla \cdot \vufp^{(0)} = 0,\\
&\parD{\theta_f^{(0)}}{t} + \textrm{Pr}_f\,\vufp^{(0)} \cdot \nabla \theta_f^{(0)} = \nabla^2 \theta_f^{(0)} - w_f^{(0)},
\end{dcases}
\end{equation*}

\begin{equation*}
\textrm{ In }\Omega_m: \hspace{.15in}
\begin{dcases}
&\vump^{(0)} = 0\, ,\\
&\varrho\, \parD{\theta_m^{(0)}}{t} =\nabla^2 \theta_m^{(0)} \hspace{.1in}(\textrm{since } \vump^{(0)} = 0),
\end{dcases}
\end{equation*}

\begin{equation*}
\mathcal{O}(1)\textrm{ On }\Gamma_i: \hspace{.15in}
\begin{dcases}
&\hat{d}\,\theta_f^{(0)} = \epsilon_T^2\,\theta_m^{(0)},\\
& \nabla \theta_f^{(0)}\cdot \mathbf{n}= \epsilon_T \, \nabla \theta_m^{(0)} \cdot \mathbf{n},\\
& \vufp^{(0)}\cdot \mathbf{n} = \hat{d}\, \vump^{(0)}\cdot \mathbf{n} \, ,\\
& 0 = u_{f,\gamma}^{(0)} \textrm{ for }\gamma=1,2,\\
&-\mathbf{n}\cdot\mathbb{T}\left(\vufp^{(0)},\pi_f^{(0)} \right)\mathbf{n}= \hat{d}^2\,\pi_m^{(0)}\, .
\end{dcases}
\end{equation*}
We notice that $\vump^{(0)}\equiv 0$ and the interface conditions reduce to 
\begin{equation*}
\textrm{On }\Gamma_i: \hspace{.15in}
\begin{dcases}
&\vufp^{(0)}=0,\\
&\hat{d} \theta_f^{(0)}  = \epsilon_T^2\, \theta_m^{(0)} ,\\
& \nabla\theta_f^{(0)}\cdot \mathbf{n}= \epsilon_T \,\nabla\theta_m^{(0)}\cdot \mathbf{n},\\
&-\mathbf{n}\cdot\mathbb{T}\left(\vufp^{(0)},\pi_f^{(0)} \right)\mathbf{n}= \hat{d}^2\,\pi_m^{(0)}\, .
\end{dcases}
\end{equation*}
The $\mathcal{O}(1)$ dynamic pressure term $\frac{1}{2}|\vufp^{(0)}\cdot \vufp^{(0)}|$ will be equal to zero at this order since $\vufp^{(0)}=0$ at the interface, and the nonlinear Lions interface condition matches its linear counterpart.

\subsection*{Balancing $\mathcal{O}(\varepsilon)$}
\begin{equation*}
\textrm{In }\Omega_f: \hspace{.15in}
\begin{dcases}
&\frac{1}{\textrm{Pr}_f}\parD{\vufp^{(1)}}{t} +\left[\vufp^{(0)}\cdot \nabla\vufp^{(1)} + \vufp^{(1)}\cdot \nabla\vufp^{(0)}\right] =\nabla^2 \vufp^{(1)} -\nabla \pi_f^{(1)} - \Raf\,\theta_f^{(1)}\, \mathbf{k},\\
&\nabla \cdot \vufp^{(1)} = 0,\\
&\parD{\theta_f^{(1)}}{t} + \textrm{Pr}_f\,\left[\vufp^{(0)}\cdot \nabla \theta_f^{(1)} + \vufp^{(1)}\cdot \nabla \theta_f^{(0)} \right]= \nabla^2 \theta_f^{(1)}  - w_f^{(1)},
\end{dcases}
\end{equation*}

\begin{equation*}
\textrm{In }\Omega_m: \hspace{.15in}
\begin{dcases}
&\vump^{(1)} =0,\\
&\varrho\, \parD{\theta_m^{(1)}}{t} =\nabla^2 \theta_m^{(1)},
\end{dcases}
\end{equation*}

\begin{equation*}
\textrm{On }\Gamma_i: \hspace{.15in}
\begin{dcases}
&\hat{d} \theta_f^{(1)}  = \epsilon_T^2\, \theta_m^{(1)} ,\\
& \nabla  \theta_f^{(1)}\cdot \mathbf{n} = \epsilon_T \,\nabla\theta_m^{(1)}\cdot \mathbf{n},\\
&\vufp^{(1)}\cdot \mathbf{n}=0,\\
&-\boldsymbol{\tau} \cdot \mathbb{T}\left(\vufp^{(0)} , \pi_m^{(0)} \right)\mathbf{n}=\alBJSJ\,u_{f,\gamma}^{(1)} \textrm{ for }\gamma=1,2,\\
&-\mathbf{n}\cdot\mathbb{T}\left(\vufp^{(1)}, \pi_f^{(1)} \right)\mathbf{n}= \hat{d}^2\,\pi_m^{(1)}\, .
\end{dcases}
\end{equation*}
The $\mathcal{O}(\varepsilon)$ dynamic pressure term $\frac{1}{2}|\vufp^{(0)}\cdot \vufp^{(1)}|$ will  equal zero at this order also (since $\vufp^{(0)}=0$ on $\Gamma_i$), and the nonlinear Lions interface condition is still equal to its linear counterpart.

\subsection*{Balancing $\mathcal{O}(\varepsilon^2)$}
\begin{equation*}
\textrm{In }\Omega_f: \hspace{.15in}
\begin{dcases}
&\frac{1}{\textrm{Pr}_f}\parD{\vufp^{(2)}}{t}+\left[ \vufp^{(1)}\cdot\nabla \vufp^{(1)} +\vufp^{(2)}\cdot\nabla \vufp^{(0)} +\vufp^{(0)}\cdot\nabla \vufp^{(2)}\right] \\
&\hspace{1in}= \nabla^2 \vufp^{(2)} -\nabla \pi_f^{(2)} - \Raf\,\theta_f^{(2)}\, \mathbf{k},\\
&\nabla \cdot \vufp^{(2)}  = 0,\\
&\parD{\theta_f^{(2)} }{t}+ \textrm{Pr}_f\,\left[ \vufp^{(1)}\cdot \nabla \theta_f^{(1)} + \vufp^{(2)}\cdot \nabla \theta_f^{(0)}+ \vufp^{(0)}\cdot \nabla \theta_f^{(2)}\right] =\nabla^2 \theta_f^{(2)}-w_f^{(2)},
\end{dcases}
\end{equation*}

\begin{equation*}
\textrm{In }\Omega_m: \hspace{.15in}
\begin{dcases}
&\frac{1}{\chi}\frac{1}{\textrm{Pr}_m}\, \parD{\vump^{(0)}}{t} + \vump^{(2)} = -\frac{1}{\hat{d}^2}\,\nabla \pi_m^{(0)} - \Raf \frac{\epsilon_T^2}{\hat{d}^4}\,\theta_m^{(0)}\,\mathbf{k},\\
&\nabla \cdot \vump^{(2)} = 0,\\
&\varrho\, \parD{\theta_m^{(2)}}{t}+\textrm{Pr}_m\,\vump^{(2)} \cdot \nabla\theta_m^{(0)}=\nabla^2 \theta_m^{(2)} - \,w_m^{(2)},
\end{dcases}
\end{equation*}

\begin{equation*}
\textrm{On }\Gamma_i: \hspace{.15in}
\begin{dcases}
&\hat{d}\theta_f^{(2)} = \epsilon_T^2\, \theta_m^{(2)}, \\
& \nabla\theta_f^{(2)}\cdot\mathbf{n}= \epsilon_T \,\nabla\theta_m^{(2)}\cdot \mathbf{n},\\
&\vufp^{(2)}\cdot\mathbf{n} = \hat{d}\,\vump^{(2)}\cdot \mathbf{n}, \\
&-\boldsymbol{\tau} \cdot \mathbb{T}\left(\vufp^{(1)} , \pi_m^{(1)} \right)\mathbf{n}=\alBJSJ\,u_{f,\gamma}^{(2)} \textrm{ for }\gamma=1,2,\\
&-\mathbf{n}\cdot \mathbb{T}\left(\vufp^{(2)}, \pi_f^{(2)} \right)\mathbf{n}= \hat{d}^2\,\pi_m^{(2)},\\
&\textrm{or }-\mathbf{n}\cdot \mathbb{T}\left(\vufp^{(2)}, \pi_f^{(2)} \right)\mathbf{n}+\frac{1}{2}\left|\vufp^{(1)}\right|^2 = \hat{d}^2\,\pi_m^{(2)}.
\end{dcases}
\end{equation*}

So, at $\mathcal{O}(\varepsilon^2)$, the dynamic pressure term finally contributes to the Lions interface condition, which does not match its linear counterpart since $\vufp^{(1)}\not\equiv 0$ on $\Gamma_i.$
The first time the dynamic pressure term influences solutions, $\vufp$ or $\vump$, is at $\mathcal{O}(\varepsilon^4)$ though. The Lions interface condition and its linear equivalent give a boundary condition for the $\pi_m^{(2)}$ term, which first shows up at $\mathcal{O}(\varepsilon^4)$ in $\Omega_m$ with Darcy's equation to solve for $\vump^{(4)}$:
\begin{align*}
\frac{1}{\chi}\frac{1}{\textrm{Pr}_m}\, \parD{\vump^{(2)}}{t} + \vump^{(4)} = -\frac{1}{\hat{d}^2}\,\nabla \pi_m^{(2)} - \Raf \frac{\epsilon_T^2}{\hat{d}^4}\,\theta_m^{(2)}\,\mathbf{k}.
\end{align*}
Thus, $\vump^{(4)}$ affects interface conditions for the $\mathcal{O}(\varepsilon^4)$ solution, $\vufp^{(4)}$.

\end{document}